\documentclass[twocolumn]{aastex701}

\newcommand\htwoco{H$_2$CO}
\usepackage[version=3]{mhchem}
\usepackage[overload]{textcase}
\usepackage{amsmath}
\usepackage{float}
\usepackage{soul}
\usepackage[font=small,labelfont=bf]{caption}
\usepackage{xcolor}
\usepackage{pifont}

\newcommand{\cmark}{\ding{51}}%
\newcommand{\xmark}{\ding{55}}%

\definecolor{MidnightBlue}{rgb}{0.0, 0.5, 0.69}
\definecolor{bittersweet}{rgb}{1.0, 0.44, 0.37}
\definecolor{green(ncs)}{rgb}{0.0, 0.62, 0.42}
\definecolor{hotmagenta}{rgb}{1.0, 0.11, 0.81}

\shorttitle{AGE-PRO:\htwoco~abundances}
\shortauthors{Chevalier et al.}
\graphicspath{{./}{figures/}}

\begin{document}

\title{The ALMA Survey of Gas Evolution of PROtoplanetary Disks (AGE-PRO): \\
Formaldehyde (\htwoco) emission and its links to disk properties}

\correspondingauthor{Ella Chevalier}
\email{ekchevalier@berkeley.edu}

\author[0009-0001-5127-7355]{Ella Chevalier}
\affiliation{Department of Astronomy, University of California, Berkeley, 
501 Campbell Hall \#3411 Berkeley, CA 94720}
\email{ekchevalier@berkeley.edu}

\author[0000-0002-0661-7517]{Ke Zhang}
\affiliation{Department of Astronomy, University of Wisconsin-Madison, 
475 N Charter St, Madison, WI 53706}
\email{ke.zhang@wisc.edu}

\author[0000-0002-4147-3846]{Miguel Vioque}
\affiliation{European Southern Observatory, Karl-Schwarzschild-Str. 2, 85748 Garching bei M\"{u}chen, Germany}
\email{miguel.vioque@eso.org}

\author[0000-0002-2358-4796]{Nicol\'{a}s T. Kurtovic}
\affiliation{Max Planck Institute for Extraterrestrial Physics, Giessenbachstrasse 1, D-85748 Garching, Germany}
\affiliation{Max-Planck-Institut fur Astronomie (MPIA), Konigstuhl 17, 69117 Heidelberg, Germany}
\email{nicokurtovic@gmail.com}

\author[0000-0001-8764-1780]{Paola Pinilla}
\affiliation{Mullard Space Science Laboratory, University College London, Holmbury St Mary, Dorking, Surrey RH5 6NT, UK}
\email{p.pinilla@ucl.ac.uk}

\author[0000-0002-1575-680X]{James Miley}
\affiliation{Joint ALMA Observatory, Alonso de C\'{o}rdova 3107, Vitacura, Santiago, Chile}
\affiliation{European Southern Observatory, Alonso de C\'{o}rdova 3107, Vitacura, Santiago, Chile}
\affiliation{Millennium Nucleus on Young Exoplanets and their Moons (YEMS), Chile}
\email{james.miley@alma.cl}

\author[0000-0003-0777-7392]{Dingshan Deng}
\affiliation{Lunar and Planetary Laboratory, the University of Arizona, Tucson, AZ 85721, USA}
\email{dingshandeng@arizona.edu}

\author[0000-0003-2251-0602]{John Carpenter}
\affiliation{Joint ALMA Observatory, Alonso de C\'{o}rdova 3107, Vitacura, Santiago, Chile}
\email{John.Carpenter@alma.cl}

\author[0000-0002-7238-2306]{Carolina Agurto-Gangas}
\affiliation{Departamento de F\'isica, Universidad T\'ecnica Federico Santa Mar\'ia, Vicu\~{n}a Mackenna 3939, San Joaqu\'in, Santiago de Chile, Chile}
\email{cagurto@das.uchile.cl}

\author[0000-0002-5991-8073]{Anibal Sierra}
\affiliation{Mullard Space Science Laboratory, University College London, Holmbury St Mary, Dorking, Surrey RH5 6NT, UK}
\affiliation{Universidad Nacional Aut\'{o}noma de M\'{e}xico, Instituto de Astronom\'{i}a, A.P. 70-264, Ciudad de M\'{e}xico 04510, M\'{e}xico}
\email{asierra@astro.unam.mx}



\begin{abstract}
Protoplanetary disks are rotating structures of gas and dust surrounding young stars, serving as the birth places of planets. Understanding the chemical evolution of organic materials in these disks is key for tracing the origins of organics in planetary systems.  Formaldehyde (\htwoco) is the most commonly detected organic molecule in protoplanetary disks. In this study, we investigate the emission of \htwoco~and its link to disk properties, using a sample of 20 Class II disks in the Lupus and Upper Sco star-forming regions spanning over 1-6\,Myr. We analyze the \htwoco~lines at 218.222 and 290.623 GHz observed as part of the AGE-PRO ALMA Large Program \citep{AGEPRO_I_overview}. Within this sample we achieve a detection rate of \htwoco~of 45\% (9/20), and set robust upper limits for the non-detections. We measure the excitation temperature and column density of the \htwoco~gas in the sources with \htwoco~detections. We combine our sample with 13 additional disks with archival \htwoco~detections and search for correlations between \htwoco~properties and disk parameters. Notably, we find strong correlations between \htwoco~line luminosity and dust radius, gas radius, dust mass, gas mass, stellar mass, and stellar luminosity. This suggests that \htwoco~emission is brighter for extended massive dust disks where \htwoco~can form via CO ice hydrogenation on grain surfaces. We find that the \htwoco~excitation temperature is also correlated with stellar mass and stellar luminosity, so more massive and luminous stars could increase \htwoco~excitation.

\end{abstract}

\keywords{Protoplanetary disks--- Astrochemistry --- Exoplanet formation --- Interferometry --- Millimeter astronomy}


\section{Introduction} \label{sec:intro}

\setcounter{footnote}{0}

The composition and potential habitability of exoplanets are strongly dependent on the processes that lead to their formation. Planets form in protoplanetary disks, which are disk-like structures of gas and dust that rotate around young stars \citep[see review by][]{williams11, manara2023}. 
As disks evolve, their chemical compositions may dramatically change over time due to the evolving radiation field, as well as the density and temperature structures \citep[e.g.,][]{Zhang20_evolution,Oberg21_review}. In particular, complex organic molecules are essential for life, and therefore understanding the origins of organic chemistry in protoplanetary disks is key for our understanding of how habitable planets form \citep{Booth2021, Krijt_PPVII,Zhang_2024RvMG, Ligterink2024}. 

Formaldehyde (\htwoco) is a small organic molecule that is considered a stepping stone to more complex organics. \htwoco~can form through two processes: (1) neutral-neutral gas-phase chemistry \citep[e.g.,][]{Fockenberg2002, Atkinson2006} and (2) grain-surface chemistry via CO ice hydrogenation \citep[e.g.,][]{Hiraoka1994, Hiraoka2002, Watanabe2002, Hidaka2004, Watanabe2004, Fuchs2009}.
Neutral-neutral gas-phase chemistry occurs more frequently in the inner disk because these reactions occur faster where the temperature and density are higher. In contrast, the CO ice hydrogenation occurs past the CO snowline, where CO freezes out onto grain surfaces. \htwoco~formed through CO ice hydrogenation is expected to be the first step towards forming oxygen-bearing, complex organic molecules (COMs). COMs are carbon-bearing molecules with six or more atoms \citep{Herbst2009}. COMs are difficult to detect in protoplanetary disks because they typically have low abundances in the gas-phase and many transitions, so the emission from each line is weaker \citep{Walsh14}. \htwoco, however, is the most commonly detected organic molecule in protoplanetary disks \citep[e.g.,][]{Pegues20}. Therefore, understanding the abundances and spatial distribution of \htwoco~in disks is the foundation for understanding the more complex organic chemistry that is difficult to observe directly. 

The distribution of \htwoco~in protoplanetary disks reveals how the two chemical formation pathways operate in different disk regions. Observations of the TW Hya disk detected two \htwoco~emission lines that provided key constraints on its spatial distribution and origins \citep{Oberg17}. The emission was found to be centrally peaked, indicating some gas-phase formation of \htwoco~in the warm inner disk, and most of emission is from the outer disk originated at intermediate heights (0.2 $<$ $z/r$ $<$ 0.4), consistent with production via CO ice hydrogenation on grain surfaces. Theoretical models similarly reproduce this pattern, with gas-phase chemistry dominating in the inner regions and grain-surface formation in the colder outer disk where CO remains frozen long enough for hydrogenation to occur \citep{Willacy_and_Woods09, Walsh14}.  However, the relative importance of these two pathways remains debated. \citet{Hernández-Vera_2024} modeled spatially resolved \htwoco~emission in HD~163296 and found that gas-phase reactions contribute significantly even in the outer disk, and
\citet{Scheltinga2021} argued that gas-phase formation could be
the dominant pathway in TW~Hya based on the emitting height and
ortho-to-para ratio of \htwoco.  

\htwoco~formation in disks could be accelerated by the presence of substructures. Substructures in disks can appear in many forms, such as axisymmetric rings and gaps or asymmetric spirals and crescents, with rings being the most common form of substructure \citep{long18, Bae2023}. These substructures can cause dust trapping by creating pressure bumps where dust accumulates as it radially drifts inward \citep{Ligterink2024}. The dust grains can be vertically transported to warmer surface layers in the disk, resulting in a good environment for the formation of organic molecules. As a result, disks with substructures can have efficient \htwoco~formation via CO ice hydrogenation on dust grains due to dust trapping. In the absence of dust traps, molecules are likely to stay trapped in ices, making them undetectable \citep{Ligterink2024}. Substructures are therefore a probable contributor to high levels of \htwoco~emission in disks.



\htwoco~has been detected in more than twenty protoplanetary disks, within both T Tauri and Herbig disks. 
\citep[e.g.,][]{Loomis15, Carney2017, Oberg17, Pegues20,Oberg21_MAPS,Booth23}. \htwoco~has also been detected in protoplanetary disks around M-stars by \citet{Pegues21}. This study analyzed five disks around M4-M5 stars, and detected \htwoco~toward one line in two disks. Both of these disks showed central depressions in the \htwoco~emission and peaks in \htwoco~emission around or beyond the edge of the pebble disk \citep{Pegues21}.   

Rotational diagram analyses have been done for various sources with multiple \htwoco~line detections, including IRS~48, HD~142527, PDS~70,
and HD~100546 \citep{vanderMarel2021, Temmink2023, Rampinelli2024,
Evans2025}. The results revealed a wide range of excitation conditions:
$T_{\rm ex}$ ranges from $\sim$20\,K in HD~142527 to over 100\,K
in IRS~48, and column densities from $2.0\times10^{13}$\,cm$^{-2}$
in PDS~70 to $2.1\times10^{14}$\,cm$^{-2}$ in HD~142527. The
inferred dominant formation pathway also varies: grain-surface
chemistry is favored in IRS~48, where a dust trap can loft
ice-coated grains to warmer layers \citep{vanderMarel2021}, while
gas-phase chemistry is favored in PDS~70 and HD~142527 - the
latter despite hosting a dust trap, because its CO snowline lies
beyond the trap \citep{Temmink2023, Rampinelli2024}.

So far, the largest population study of \htwoco~in protoplanetary disks was by \citet{Pegues20}, which used a sample of 15 disks with stellar mass between 0.5-2.0 M$_{\odot}$ and age between 1-15 Myr. The 15 disks are T Tauri and Herbig Ae disks, with six disks in the Taurus-Auriga region, five disks in the Upper Sco region, one disk in the Lupus region, and one disk in the Ophiuchus region. Five of these disks (IM Lup, GM Aur, AS 209, HD 163296, and MWC 480) are analyzed in more detail in \citep{Oberg21_MAPS}. With this sample, \citet{Pegues20} detected \htwoco~in 13/15 disks and tentatively in one additional disk. These detections showed diverse emission morphologies, including centrally peaked, centrally depressed, and ring-like structures. In the four disks with detections toward multiple \htwoco~lines, they found excitation temperatures between 11-37 K and column densities between 1.9-21$\times$10$^{12}$ cm$^{-2}$. However, this sample of 15 disks was inhomogeneous, with a wide range of stellar masses ($\sim0.5-2.0M_\odot$) and an uneven age distribution (e.g., most of the old sources are the sources with the highest stellar masses). As a result, no conclusions were drawn about the dependence between \htwoco~line flux and disk and stellar properties. In this paper, we use a more homogeneous sample with disks from the AGE-PRO sample \citep{AGEPRO_I_overview} that have a narrow range of stellar masses but a wide span of ages. The AGE-PRO sample also allows us to test dependencies with gas mass for the first time. 
We aim to evaluate how \htwoco~emission is connected to disk properties.

This paper presents a detailed study of the \htwoco~emission in 20 Class II disks around 0.3-0.8\,M$_\odot$ stars and tests how \htwoco~emission changes with disk parameters such as gas mass, gas radius, dust mass, dust radius, stellar luminosity, and stellar mass. \S 2 describes our disk sample, the observations, and measurements of line fluxes, excitation temperature, and total column density. \S 3 presents our results on detection rate, line fluxes, and correlations with disk properties. In \S 4, we discuss potential explanations for the correlations we find with disk properties, including the implications on \htwoco~evolution and chemistry. \S 5 summarizes our findings and discusses future research directions for organics in protoplanetary disks.

\section{Method} \label{sec:age-pro design}
\subsection{Disk Sample and Data calibration}

Our sample composites of 20 Class II disks which were part of the the ALMA Survey of Gas Evolution of PROtoplanetary Disks (AGE-PRO), a Cycle 8 Large Program of the Atacama Large Millimeter/submillimeter Array (ALMA) \citep{AGEPRO_I_overview}. The AGE-PRO program was designed to trace the evolution of gas disk masses and sizes, which selected 30 disks from three nearby star-forming regions of different ages: Ophiuchus (embedded disks, 0.5-1\,Myr), Lupus (middle age, 1-3\,Myr), and Upper Sco (the end of gas disk lifetime, 2-6\,Myr) \citep{AGEPRO_I_overview}. This work focuses on the 20 Class II disks from the Lupus and Upper Sco regions. See Table~\ref{table:stars} for the stellar properties of the 20 sources.

The AGE-PRO sample was selected with four main criteria: (1) sources with known stellar spectral type between M3-K6, roughly corresponding to a stellar mass (M$_\star$) of 0.3-0.8\,M$_\odot$, see Table~\ref{table:stars}. (2) sources without known companions or that are in wide-separation binaries ($>$600\,AU), as close binaries may evolve differently due to tidal interactions. (3)  
For Lupus and Upper Sco regions, Class II sources were selected to exclude debris disks. (4) sources were selected from previous detections of mm continuum and CO line emission. Ten disks were selected from each region that covers the spread of continuum luminosities in the region \citep{AGE-PRO_II_Ophiuchus,Deng_Lupus,AGE-PRO_IV_UpperSco}.

The analysis presented here is based on the AGE-PRO fiducial images,  for two \htwoco~lines in Bands 6 and 7.  The angular resolution is $\sim0\farcs35 $ for Band 6 and $\sim0\farcs70 $ for Band 7 \citep{Deng_Lupus}, and the velocity resolution is 0.2 and 1.552\,km/s for \htwoco~in Band 6 for Lupus and Upper Sco sources, respectively and 1.164 km/s for \htwoco~in Band 7 \citep{Deng_Lupus,AGE-PRO_IV_UpperSco}. Further information including the observational logs, line frequencies, and full details of the data calibration and imaging are described in  \citet{AGEPRO_I_overview, Deng_Lupus, AGE-PRO_IV_UpperSco}. Two \htwoco~lines are covered by our spectral setups: p-3$_{03}$-2$_{02}$ at 218.222 GHz and p-4$_{04}$-3$_{03}$ at 290.623 GHz (in this paper, we abbreviate these lines as p-3-2 and p-4-3, respectively). The line information is shown in Table~\ref{tab:line_properties}.

\begin{deluxetable*}{lccccclcccccc}\tablewidth{\textwidth}
\tabletypesize{\scriptsize}
\tablecaption{Host Star Properties of the sample \label{table:stars}}
\tablewidth{0pt}
\tablehead{
\colhead{Index} & \colhead{Source Name} & \colhead{RA} & \colhead{Dec}  & \colhead{Dist} & \colhead{Class} & 
\colhead{SpT} & \colhead{$T_{\rm eff}$} & \colhead{$L_\ast$} & \colhead{Av} & \colhead{$M_\ast$} & \colhead{$\log{\dot{M}_\ast}$} & \colhead{Refs.} 
\\
\colhead{} & \colhead{} &  \colhead{(h m s)} & \colhead{(d m s)} & \colhead{(pc)} &  & & 
\colhead{(K)} & \colhead{($L_\odot$)} & \colhead{(mag)}&\colhead{($M_\odot$)}  & \colhead{($M_\odot$\, ${\rm yr}^{-1}$)} & \colhead{}
}
\decimalcolnumbers
\startdata
 Lupus 1     &  Sz65                     &  15:39:27.753    &  -34:46:17.577   &  153.0&  II    &  K7   &    4060&   0.87&    0.6&   0.68&    -9.5& 1,2,3\\
 Lupus 2     &  Sz71                     &  15:46:44.709    &  -34:30:36.054   &  154.1&  II    &  M1.5 &    3632&   0.33&    0.5&   0.42&    -9.0& 1,2,3\\
 Lupus 3     &  J16124373-3815031        &  16:12:43.736    &  -38:15:03.471   &  158.7&  II    &  M1   &    3705&   0.39&    0.8&   0.48&    -9.1& 1,2,3\\
 Lupus 4     &  Sz72                     &  15:47:50.608    &  -35:28:35.779   &  155.5&  II    &  M2   &    3560&   0.27&    0.8&   0.39&    -8.6& 1,2,3\\
 Lupus 5     &  Sz77                     &  15:51:46.941    &  -35:56:44.531   &  154.8&  II    &  K7   &    4060&   0.59&    0.0&   0.73&    -8.7& 1,2,3\\
 Lupus 6     &  J16085324-3914401        &  16:08:53.227    &  -39:14:40.553   &  161.5&  II    &  M3   &    3415&   0.20&    1.9&   0.31&   -10.0& 1,2,3\\
 Lupus 7     &  Sz131                    &  16:00:49.414    &  -41:30:04.263   &  159.1&  II    &  M3   &    3415&   0.15&    1.3&   0.31&    -9.2& 1,2,3\\
 Lupus 8     &  Sz66                     &  15:39:28.264    &  -34:46:18.450   &  154.4&  II    &  M3   &    3415&   0.22&    1.0&   0.30&    -8.5& 1,2,3\\
 Lupus 9     &  Sz95                     &  16:07:52.293    &  -38:58:06.446   &  159.2&  II    &  M3   &    3415&   0.27&    0.8&   0.30&    -9.4& 1,2,3\\
 Lupus 10    &  V1094Sco                 &  16:08:36.160    &  -39:23:02.879   &  154.8&  II    &  K6   &    4205&   1.21&    1.7&   0.82&    -7.9& 1,2,3\\
 UppSco 1 &  J16120668-3010270        &  16:12:06.664    &  -30:10:27.617   &  131.9&  II    &  M0.5 &    3700&   0.25&    0.1&   0.51&    -9.4& 4-9\\
 UppSco 2 &  J16054540-2023088        &  16:05:45.379    &  -20:23:09.330   &  137.6&  II    &  M4.5 &    3020&   0.07&    0.3&   0.13&    -9.4& 4-9\\
 UppSco 3 &  J16020757-2257467        &  16:02:07.556    &  -22:57:47.424   &  139.6&  II    &  M2   &    3490&   0.15&    0.5&   0.37&   -11.0& 4-9\\
 UppSco 4 &  J16111742-1918285        &  16:11:17.406    &  -19:18:29.231   &  136.9&  II    &  M0.25&    3735&   0.35&    0.9&   0.50&     \nodata& 4-9\\
 UppSco 5 &  J16145026-2332397        &  16:14:50.249    &  -23:32:40.238   &  144.0&  II    &  M3   &    3360&   0.11&    1.4&   0.29&     \nodata& 4-9\\
 UppSco 6 &  J16163345-2521505        &  16:16:33.429    &  -25:21:51.163   &  158.4&  II    &  M0.5 &    3700&   0.18&    1.1&   0.52&   -10.9& 4-9\\
 UppSco 7 &  J16202863-2442087        &  16:20:28.622    &  -24:42:09.174   &  152.7&  II    &  M2   &    3490&   0.23&    1.7&   0.34&     \nodata& 4-9\\
 UppSco 8 &  J16221532-2511349        &  16:22:15.324    &  -25:11:35.672   &  139.0&  II    &  M3   &    3360&   0.14&    1.9&   0.29&     \nodata& 4-9\\
 UppSco 9 &  J16082324-1930009        &  16:08:23.247    &  -19:30:00.980   &  137.0&  II    &  M0   &    3880&   0.24&    0.9&   0.56&    -9.1& 4-9\\
 UppSco 10&  J16090075-1908526        &  16:09:00.739    &  -19:08:53.284   &  136.9&  II    &  M0   &    3630&   0.35&    1.2&   0.53&    -8.8& 4-9\\
 \enddata
\tablecomments{Col.~(1) AGE-PRO index (2) Source name,
(3) ICRS RA, (4) Dec based on 1.3\,mm continuum center in ALMA images, (5) Distance, Col.~(6) Disk class, (7) Spectral type, (8) Stellar effective temperature, (9) Stellar luminosity, (10) V band extinction magnitude, (11) Stellar mass, (12) Mass accretion rate, (13) References.}
\tablerefs{All distances are adopted from geometric distance of \citet{BailerJones2021}. In Col.~(13), the references for Class, Spectral type, \{$T_{\rm eff}$, $L_\ast$\}, A$_v$, and accretion rates:   1 = \citet{Alcala14_Lupus}, 2 = \citet{Alcala17_Lupus}, 3 = \citet{Deng_Lupus}, 4 = \citet{Luhman2022_uppstars}, 5 = \citet{Manara2020}, 6 = \citet{Carpenter_2025}, 7 = \citet{Manara_PPVII}, 8 = \citet{Fang2023}, 9 = \citet{AGE-PRO_IV_UpperSco}.}
\end{deluxetable*}

\begin{table}
\begin{center}

\scalebox{0.9}{

\hspace{-2cm}\begin{tabular}{ ccccc } 
     \hline
     \hline
     H$_2$CO Line & Frequency & $A_{ul}$  & $E_u$ & $g_u$ \\ 
     & (GHz) & ($10^{-4}s^{-1}$) & (K) & \\
     \hline
     p-3$_{03}$-2$_{02}$  & 218.222 & 2.818 & 20.956 & 7 \\ 
    p-4$_{04}$-3$_{03}$ & 290.623 & 6.903 & 34.904 & 9 \\
    \hline
    \end{tabular} 
}
\caption{The line properties for the two \htwoco~lines detected in the AGE-PRO disks. All values in this table are from the Cologne Database for Molecular Spectroscopy \citep{ENDRES201695}.}
\label{tab:line_properties}
\end{center}
\end{table}

\subsection{Analysis}


Here we describe the methods used to measure the \htwoco~line fluxes and create the radial profiles. We first generate moment-zero maps for the \htwoco~lines of 20 disks by applying Keplerian masks to channel maps. For the Lupus sources, we use the Keplerian mask parameters for individual disks listed in \cite{Deng_Lupus}. For the Upper Sco sources, we generate Keplerian masks based on the inclination and position angles measured from 1.3\,mm continuum emission by \cite{ageproX}. The best-fitting masks are decided by eye evaluation of masks in individual channel maps. For sources without $3\sigma$ \htwoco~line detections, we generate masks based on $^{12}$CO $J$=2-1 line emission. The Keplerian mask parameters for the Lupus and Upper Sco disks are listed in Appendix Table~\ref{tab:mask_params}. Our line fluxes for the Upper Sco disks are consistent with those obtained using the Keplerian mask parameters from \cite{AGE-PRO_IV_UpperSco}, but we use a slightly larger Keplerian mask to be more conservative for the fainter Upper Sco disks.

\begin{figure*}[!ht]
\centering
\vspace{-0.cm}
\includegraphics[width=\textwidth]{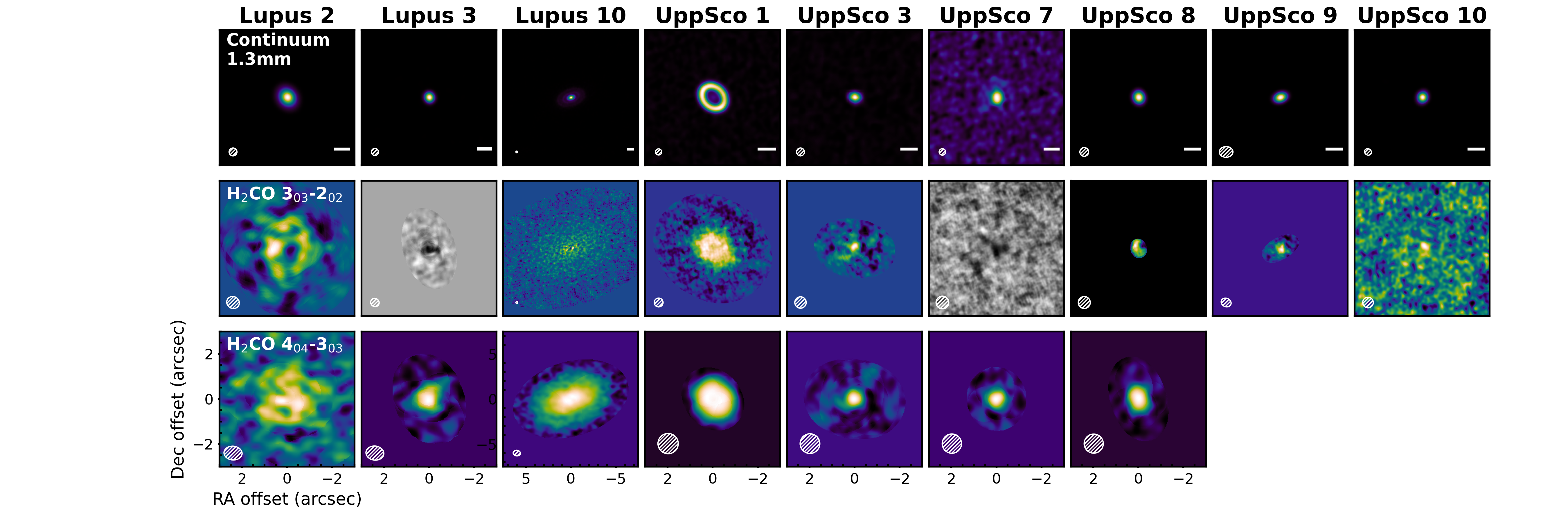}
\caption{Moment-zero maps for the Lupus and Upper Sco sources with \htwoco~detections. The first row for each region shows the total continuum emission at 1.3mm. The remaining two rows show the total emission for \htwoco~in both wavelength bands. The beam size is shown in the bottom left. The grayscale panels represent the lines without confident 3$\sigma$ \htwoco~detections. No data is available for the \htwoco~p-4-3 line for Upper Sco 9 and 10. The moment-zero maps of all Lupus and Upper Sco sources can be found in Appendix Figure~\ref{fig:final_m0_all}.}
\label{fig:total_m0}

\end{figure*}

\begin{figure*}[!t]
\centering
\vspace{-0.cm}
\includegraphics[width=\textwidth]{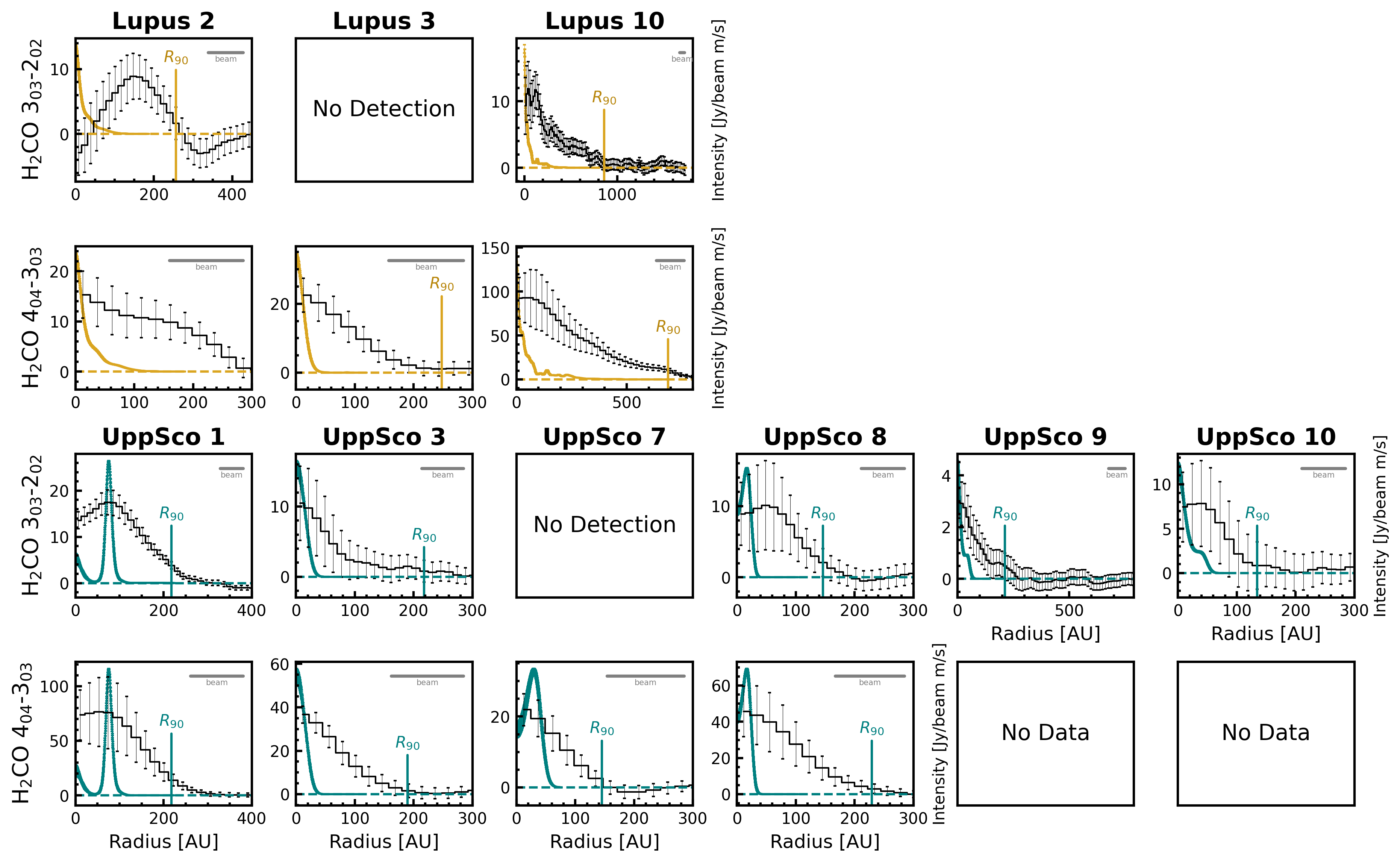}
\vspace{-0.cm}
\caption{Radial profiles for the Lupus and Upper Sco sources with \htwoco~detections. The R$_{90}$ value, or the radius at which 90$\%$ of the flux is enclosed, is labeled. The beam size is shown in the top right corner. The dust radial profiles are overplotted in gold and teal for Lupus and Upper Sco, respectively, but are not shown to scale. While Lupus 3 is less resolved  (larger beam sizes) compared to Lupus 10, clear \htwoco~detections are seen in both the moment-zero maps and radial profiles. Upper Sco 3, 7, and 8 are also less resolved but still have clear \htwoco~detections in both the moment-zero maps and radial profiles.}
\label{fig:total_rp}
\vspace{-0.cm}
\end{figure*}

After generating the moment-zero maps, we use the curve-of-growth method to estimate the line fluxes and uncertainties of \htwoco~line emission \citep{AGEPRO_I_overview,Deng_Lupus}. We take line fluxes for the Lupus disks from \citet{Deng_Lupus}, and calculate the line fluxes for Upper Sco following the same methodology. Basically, elliptical apertures are projected onto the moment-zero maps, based on the inclination and position angle estimated from continuum emission \citep{ageproX}. The 100\% flux radius, denoted as R$_{100}$, is the radius corresponding to the first maxima in the integrated flux, and the total flux is measured within this radius. Because the Band~6 and Band~7 data have different angular
resolutions ($\sim$0\farcs35 vs.\ $\sim$0\farcs70), R$_{100}$
is determined independently for each \htwoco~transition, and
fluxes are integrated within the corresponding aperture. The resulting \htwoco~line fluxes and their uncertainties are given in Table~\ref{tab:flux}. 


\begin{table*}
\hspace*{-0.5cm}\begin{tabular}{ lllllll } 
 \hline
 \hline
 Disk & H$_2$CO Line & Flux & Peak Flux & Velocity Range & rms & Beam Size \\ 
  &  & (mJy km s$^{-1}$) & (mJy beam$^{-1}$  & (km s$^{-1}$) & (mJy  & (") \\ 
  &  & &  km s$^{-1}$) &     & beam$^{-1}$) & \\ 
 \hline
 Lupus 1 & p-3-2 & $<$5.5 & 13.1  &  -1-9 & 1.7 & 0.43 $\times$ 0.30 \\ 
 ... & p-4-3 & $<$29.5 & 2.5 &  -1-9 & 1.0 & 0.81 $\times$ 0.61 \\ 
 Lupus 2 &  p-3-2 & 96.2 $\pm$ 7.6 & 52.5 &  -3-12 & 2.3 & 0.59 $\times$ 0.59 \\ 
 ... &  p-4-3 & 145.8 $\pm$ 12.5 & 45.3 &  -3-12 & 1.0 & 0.80 $\times$ 0.61  \\ 
 Lupus 3 & p-3-2 & $<$30.8 & 9.8 &  -2-11 & 1.0 & 0.80 $\times$ 0.63 \\ 
 ... & p-4-3 & 44.1 $\pm$ 8.5 & 10.9 &  -2-11 & 1.0 & 0.80 $\times$ 0.63  \\ 
 Lupus 4 & p-3-2 & $<$4.5 & 5.5 &  -1-9 & 2.1 & 0.39 $\times$ 0.35 \\
 ... & p-4-3 & $<$19.1 & 2.7 &  -1-9 & 1.0 & 0.80 $\times$ 0.62\\ 
 Lupus 5 & p-3-2 & $<$19.4 & 7.2 &  -1-9 & 2.0 & 0.36 $\times$ 0.33 \\
  ... & p-4-3 & $<$20.1 & 1.0 &  -1-9 & 1.0 & 0.81 $\times$ 0.61 \\
 Lupus 6 & p-3-2 & $<$31.7 & 15.6 &  -1-9 & 4.0 & 0.43 $\times$ 0.41 \\
 Lupus 7 & p-3-2 & $<$11.9 & 35.9  &  -1-9 & 4.0 & 0.45 $\times$ 0.40 \\
 ... & p-4-3 & $<$7.9 & 9.9 &  -3-12 & 1.0 & 0.80 $\times$ 0.64  \\ 
 Lupus 8 & p-3-2 & $<$11.1 & 3.6 &  -1-9 & 4.5 & 0.40 $\times$ 0.35 \\
  ... & p-4-3 & $<$14.7 & 1.2 &  -1-9 & 1.9 & 0.81 $\times$ 0.61 \\
 Lupus 9 & p-3-2 & $<$19.5 & 14.3 &  -1-9 & 3.9 & 0.45 $\times$ 0.43 \\
  ... & p-4-3 & $<$12.6 & 2.9 &  -1-9 & 1.0 & 0.79 $\times$ 0.64 \\
 Lupus 10 & p-3-2 & 1916.7 $\pm$ 10.0 & 928.7 &  0-10 & 3.2 & 0.31 $\times$ 0.31 \\ 
 ... &  p-4-3 & 2269.4 $\pm$ 26.9 & 519.3  &  -2-12 & 2.1 & 0.80 $\times$ 0.63  \\ 
 UpperSco 1 & p-3-2 & 517.5 $\pm$ 4.9 & 114.2 &  -3-12 & 0.6 & 0.39 $\times$ 0.39  \\ 
  ... &  p-4-3 & 416.8 $\pm$ 4.9 & 101.9 &  -3-12 & 1.0 & 0.70 $\times$ 0.58  \\
 UpperSco 2 & p-3-2 & $<$5.9 & 3.5 &  -5-15 & 0.4 & 0.46 $\times$ 0.46\\
  ... & p-4-3 & $<$10.5 & 2.2  &  -2-12 & 0.7 & 0.91 $\times$ 0.91 \\
 UpperSco 3 & p-3-2 & 37.5 $\pm$ 4.4 & 5.1 &  -3-12 & 0.4 & 0.50 $\times$ 0.50 \\ 
 ... &  p-4-3 & 40.5 $\pm$ 5.9 & 5.5 &  -3-12 & 0.5 & 0.81 $\times$ 0.60 \\ 
 UpperSco 4 & p-3-2 & $<$6.3 & 0.3 &  -3-12 & 0.4 & 0.49 $\times$ 0.49 \\
  ... & p-4-3 & $<$15.6 & 6.4 &  -3-12 & 0.7 & 0.85 $\times$ 0.85 \\
 UpperSco 5 & p-3-2 & $<$7.2 & 15.1 &  -2-12 & 2.0 & 0.36 $\times$ 0.33\\
  ... & p-4-3 & $<$8.9 & 2.0 &  -2-14 & 0.5 & 0.86 $\times$ 0.86\\
 UpperSco 7 & p-3-2 & $<$10.7 & 0.4 &  -2-12 & 0.5 & 0.56 $\times$ 0.56\\ 
 ... & p-4-3 & 30.3 $\pm$ 3.3 & 8.1 &  -2-12 & 0.7 & 0.86 $\times$ 0.86\\ 
 UpperSco 8 & p-3-2 & 41.2 $\pm$ 3.6 & 12.1 &  -3-12 & 0.5 & 0.35 $\times$ 0.30 \\ 
 ... &  p-4-3 & 99.6 $\pm$ 4.0 & 15.2 &  -3-12 & 0.8 & 0.65 $\times$ 0.49 \\ 
 UpperSco 9& p-3-2 & 13.6 $\pm$ 0.9 & 1.8 &  -3-12 & 0.1 & 0.45 $\times$ 0.37 \\ 
 UpperSco 10 & p-3-2 & 38.6 $\pm$ 5.2 & 6.9 &  -4-12 & 0.6 & 0.47 $\times$ 0.47\\ 
\hline
\end{tabular} \\
\caption{Line fluxes of the two \htwoco~lines for the 20 disks. The 3$\sigma$ upper limits are calculated for the sources without clear detections. The peak flux in column 2 is the maximum value in the \htwoco~emission spectrum, and the uncertainty is the noise of the spectrum corresponding to the peak value, only including the statistical uncertainty without the systematic flux calibration uncertainty. We do not have p-4-3 line fluxes for Upper Sco 9 and 10 because archival observations did not cover the p-4-3 line.}
\label{tab:flux}
\end{table*}

We use the \texttt{GoFish} package \citep{GoFish} to compute the radial profiles for the \htwoco~line emission using the \texttt{radial$\_$profile()} function. We use the position and inclination angles given in \citet{ageproX} (see Table \ref{tab:mask_params}) to convert image coordinates to disk-plane coordinates. We use a bin width \texttt{dr} that is four times the pixel size ($0\farcs025$ for Band 6 and $0\farcs04$ for Band 7 (\cite{AGEPRO_I_overview}), and we set the minimum radius to zero and the maximum radius to five times the $R_{\text{max}}$ value given in Table \ref{tab:mask_params} under the ``CO Mask Radius" column. The uncertainty is calculated by taking the standard deviation within each radial bin.

Figure~\ref{fig:total_m0} shows the moment-zero maps of the four Lupus disks and five Upper Sco disks with significant \htwoco~detections. All moment-zero maps are shown in Appendix Figure~\ref{fig:final_m0_all}. The radial profiles are shown in Figure~\ref{fig:total_rp}.

\subsection{Rotational diagram method}



 When two or more \htwoco~lines are detected toward a disk, the rotational diagram method can be used to determine the excitation temperatures and column densities \citep{Goldsmith1999}. Following the approach of \cite{Loomis2018}, the excitation temperature $T_{\text{ex}}$ and total column density $N_{\text{tot}}$ can be related to the upper level column density $N_u$ and the upper level degeneracy $g_u$,
\begin{equation}
    \ln\frac{N_u}{g_u}=\ln N_{\text{tot}} - \ln Q(T_{\text{ex}})-\frac{E_u}{kT_{\text{ex}}} \label{eq:rot_diag}
\end{equation}
where $Q(T_{\text{ex}})$ is the partition function. In an optically thin case, the upper level column density can be calculated from the integrated flux density $S_\nu \Delta\nu$,
\begin{equation}
    N_u^{thin}=\frac{4\pi S_\nu \Delta\nu}{A_{ul}\Omega hc}
\end{equation}
where $\Omega$ is the solid angle of the area over which the flux is integrated. The solid angle is found by converting the R$_{90}$ radius, at which 90$\%$ of the flux is enclosed, to radians. We use R$_{90}$ rather than R$_{100}$ because \htwoco~is not uniformly distributed and does not have high signal-to-noise, so R$_{90}$ is usually more robust than R$_{100}$. $A_{ul}$ is the Einstein coefficient, $h$ is the Planck constant, and $c$ is the speed of light. 

This derivation assumes that the line is optically thin. To correct for optical depth, the upper level column density $N_u^{\text{thin}}$ must be multiplied by the optical depth correction factor $C_{\tau ul}$.
\begin{equation}
    N_u=C_{\tau ul}N_u^{\text{thin}} \label{eq:Nu}
\end{equation}
\begin{equation}
    C_{\tau ul}=\frac{\tau_{ul}}{1-e^{-\tau_{ul}}}
\end{equation}
Following \cite{Pegues20}, the optical depth $\tau_{ul}$ can be calculated by Taylor expansion (assuming the line is optically thin, we expand around $\tau_{ul}=0$).

\begin{align}
    \tau_{ul} &= \left ( \frac{A_{ul}c^3}{8\pi \nu^3 \Delta \nu} \right ) N_u(e^{h\nu/(k_B T_{ex})}-1) \\ 
    \tau_{ul}-\frac{\tau_{ul}^2}{2}+\frac{\tau_{ul}^3}{6}&\approx \left ( \frac{A_{ul}c^3}{8\pi \nu^3 \Delta \nu} \right ) N_u^{\text{thin}}(e^{h\nu/(k_B T_{ex})}-1) \label{eq:tau}
\end{align}

Once the optical depth correction is applied, the rotational diagram is created using Equation~\ref{eq:rot_diag} and Equation~\ref{eq:Nu}. 
\begin{equation}
    \ln\frac{N_u}{g_u}+\ln C_{\tau_{ul}}=\ln N_{\text{tot}} - \ln Q(T_{\text{ex}})-\frac{E_u}{kT_{\text{ex}}} \label{eq:new_rd}
\end{equation}
The optical depth is recalculated using Equation~\ref{eq:new_rd}. This process is repeated until the difference in optical depth between iterations is less than 0.001$\%$ or until 5000 iterations has passed. The optical depth converges before 5000 iterations for all disks. At each iteration, we calculate the slope m and the y-intercept b from a linear fit. We then use Markov chain Monte Carlo (MCMC) sampling with \texttt{emcee.EnsembleSampler} to explore the probability distribution of the parameters with 4 ensemble walkers and an MCMC chain with 5000 steps. From the resulting MCMC chains, flat samples for m and b are created by discarding the first 100 steps and thinned by taking every 15th step. 

\begin{figure*}[!ht]
\centering
\vspace{-0.cm}
\scalebox{1}{
\includegraphics[width=1.0\textwidth]{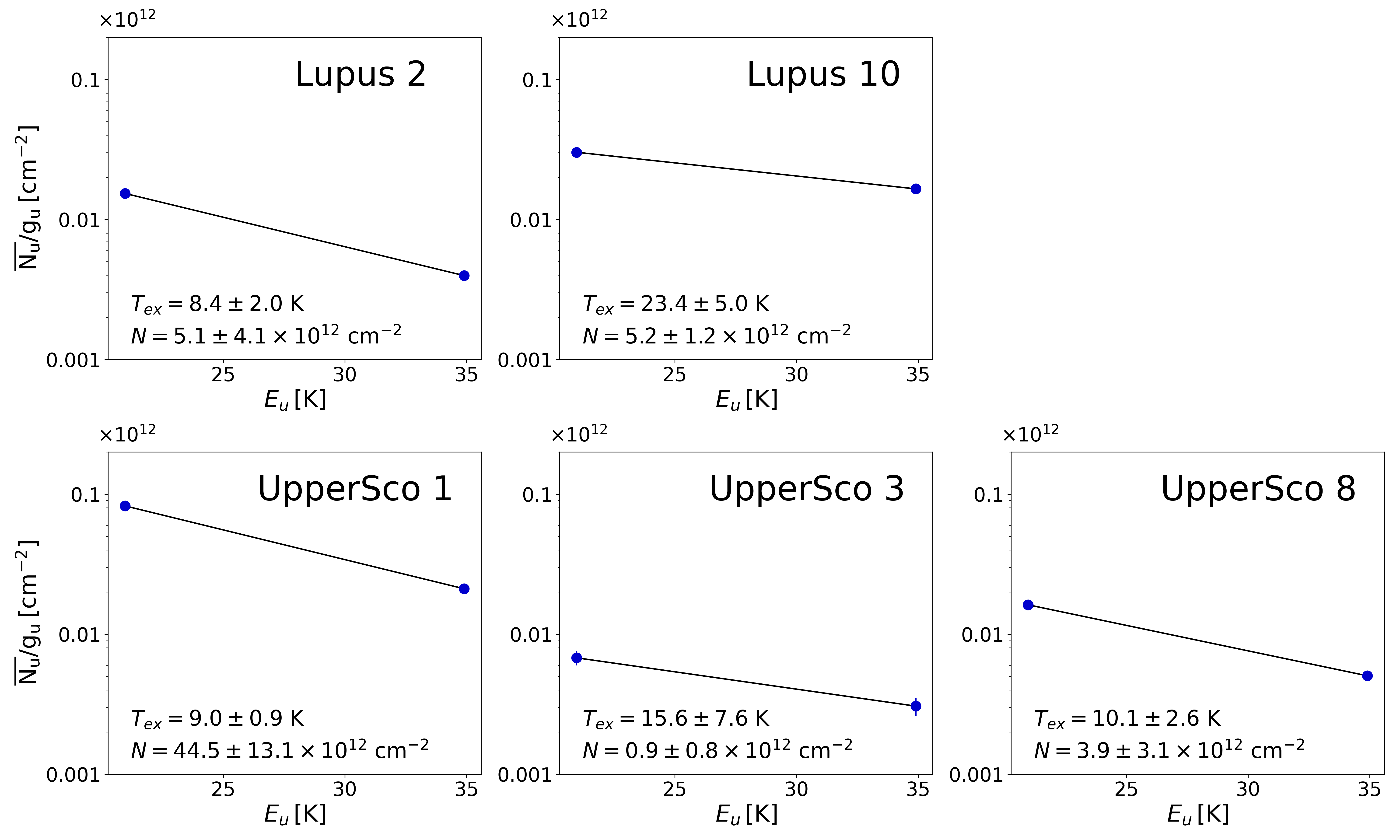}
}
\vspace{-0.cm}
\caption{The rotational diagrams for the five sources with two \htwoco~detections. The diagram shows the linear fit between the two data points. From the linear fit, the slope is used to calculate the excitation temperature and the y-intercept is used to calculate the total column density. The errorbars are too small to be visible.}
\label{fig:rot_diag}
\vspace{-0.cm}
\end{figure*}

The final m and b values are the means of the resulting flat sample arrays, and the errors are the standard deviations. $T_{\text{ex}}$ is calculated as $T_{\text{ex}} = -\frac{1}{m}$, and the error for $T_{\text{ex}}$ is calculated by $\delta T_{\text{ex}} = T_{\text{ex}} \|\frac{\delta m}{m}\|$. The total column density is calculated as $N_u^{\text{thin}}=Qe^b$. The partition function Q is interpolated using values from the CDMS catalog \citep{MULLER2001, MULLER2005215, ENDRES201695}. We use the total \htwoco~partition function that assumes nuclear spin-weights of 3 and 1 for the ortho-\htwoco~and para-\htwoco, respectively. The error for $N_u^{\text{thin}}$ is calculated by $\delta N_u^{\text{thin}} = N_u^{\text{thin}} \delta b$. Once the optical depth converges, the final $T_{\text{ex}}$ and $N_{\text{tot}}$ values have been found and the optical depth is calculated with Equation~\ref{eq:tau}. We calculate the error in optical depth by taking the standard deviation of a set of 5000 optical depth calculations made by choosing the final $T_{\text{ex}}$ and $N_u^{\text{thin}}$ randomly from a Gaussian distribution with a mean of the calculated value $T_{\text{ex}}$ or $N_u^{\text{thin}}$ and a standard deviation of the calculated error $\delta T_{\text{ex}}$ or $\delta N_u^{\text{thin}}$. The final excitation temperatures, column densities, and optical depths are shown in Table~\ref{tab:temp_and_cd}, and Figure~\ref{fig:rot_diag} shows the rotational diagrams for the five disks with two \htwoco~detections.

Since we only have two \htwoco~lines, we expect higher uncertainties on our excitation temperatures than other studies that have detections toward more \htwoco~lines. The more \htwoco~lines detected, the higher the precision that can be reached with the rotational diagram method. As a result, it is likely that our uncertainty in $T_{ex}$ and $N_u^\text{thin}$ is underestimated. Additionally, in this work we only include the noise level in the channel maps when determining the flux uncertainty. The absolute flux calibration uncertainty of our ALMA observations is around 5-10\%, which adds additional uncertainty to the flux and consequently to $T_{\text{ex}}$ and $N_u^{\text{thin}}$. We also assume local thermodynamic equilibrium (LTE) conditions, which may not be the case. With only two points in our rotational diagram, it is unclear how representative these points are of the total population distribution. In our sample, Upper Sco 1 has a particularly low uncertainty in $T_{\text{ex}}$. This is likely because Upper Sco 1 has large line fluxes and an SNR that is ten times higher than the other sources. The uncertainty in flux for Upper Sco 1 is relatively small compared to the strength of the line flux, resulting in a similarly low uncertainty in $T_{\text{ex}}$.

\begin{table*}
\hspace{1.75cm}\begin{tabular}{ llll } 
 \hline
 \hline
 Disk & $T_{\text{ex}}$ (K)  & $N_{\text{tot}}$ ($10^{12}$ cm$^{-2}$)& $\tau_{\text{ul}}$  \\ 
 \hline
Lupus 2 &  8.4 $\pm$ 2.0 & 5.1 $\pm$ 4.1 & 1.78 ($3_{03}-2_{02}$), 1.02 ($4_{04}-3_{03}$)\\ 
Lupus 10 &  23.4 $\pm$ 5.0 & 5.2 $\pm$ 1.2 & 0.36 ($3_{03}-2_{02}$), 0.38 ($4_{04}-3_{03}$)\\ 
UpperSco 1 &  9.0 $\pm$ 0.9 & 44.5 $\pm$ 13.1 & 3.66 ($3_{03}-2_{02}$), 2.98 ($4_{04}-3_{03}$)\\ 
UpperSco 3 &  15.6 $\pm$ 7.6 & 0.9 $\pm$ 0.8 & 0.14 ($3_{03}-2_{02}$), 0.13 ($4_{04}-3_{03}$)\\
UpperSco 8 &  10.1 $\pm$ 2.6 & 3.9 $\pm$ 3.1 & 1.25 ($3_{03}-2_{02}$), 0.75 ($4_{04}-3_{03}$)\\
\hline
\end{tabular} \\
\caption{The excitation temperature and total column density are derived iteratively using linear fits of Figure~\ref{fig:rot_diag}, applying the optical depth correction at each step until the optical depth converges or 5000 iterations are reached. The optical depth converges before 5000 iterations for all disks.}
\label{tab:temp_and_cd}
\end{table*}

For the four disks with only one \htwoco~line detected, we constrain the column density with two fixed excitation temperatures at 10 and 40\,K. These temperatures span the range of excitation temperatures for \htwoco~reported in literature \citep[e.g.][]{Qi13, Pegues20}. We then report a range for the total column density.
The rotational diagram method is also used for these disks, except the slope is fixed. The optical depth and the optical depth correction factor are calculated for both excitation temperatures of 10 and 40\,K. The total column density is recalculated using the optical depth correction factor. Since the excitation temperature is fixed, no iteration on the optical depth is needed.

\section{Results}

\subsection{\htwoco~detections}

Both \htwoco~lines are detected at the 3$\sigma$ level in integrated fluxes in five disks (Lupus 2 and 10, and Upper Sco 1, 3, and 8). Only \htwoco~line p-3-2 is detected in Upper Sco 9 and Upper Sco 10, but no \htwoco~p-4-3 observations were taken for these two disks. Only \htwoco~line p-4-3 is detected in Lupus 3 and Upper Sco 7. In total, there are five disks with two \htwoco~detections and four disks with one \htwoco~detection. This means that we recover a detection rate for the p-3-2 line of 20\% for Lupus and 50\% for Upper Sco, and a detection rate for the p-4-3 line of 30\% in Lupus and and 40\% in Upper Sco. A significant detection requires both a 3$\sigma$ line flux measurement and visible emission in the channel maps. Very faint \htwoco~emission is seen toward Upper Sco 2 and Upper Sco 4 for both lines, toward Lupus 3 and Upper Sco 7 for \htwoco~line p-3-2, and toward Lupus 7 for the p-4-3 line. However, the signal-to-noise ratios are not sufficient for 3$\sigma$ detections. The total line fluxes and upper limits for non-detections are displayed in Table~\ref{tab:flux}. The line fluxes vary by more than two orders of magnitude: ranging from 13.6 mJy km s$^{-1}$ in Lupus 7 to 2269.4 mJy km s$^{-1}$ in Lupus 10 for line p-4-3. Lupus 10 has the largest line fluxes toward both lines. However, Lupus 10 is a very large disk that is not representative of a typical disk \citep{ageproXI}. With the exception of Upper Sco 1, the line flux is larger for p-4-3 line than for the p-3-2 line for all disks with two detections. 



Looking at the radial profiles in Figure~\ref{fig:total_rp}, \htwoco~emission generally peaks towards the center of the disk and decreases radially outward. For the p-4-3 line, the beam size is large ($\sim$ 0\farcs8) and therefore the radial profiles are not well resolved. The p-3-2 line observations have smaller beam sizes ($\sim$ 0\farcs3) and thus we focus on the radial features seen in the p-3-2 lines. Notably, Lupus 2, Upper Sco 1, and Upper Sco 8 show a central depression. This could be indicative of an inner cavity in the disk, although while Upper Sco 1 shows an inner cavity in the CO emission, Lupus 2 and Upper Sco 8 do not \citep{AGE-PRO_IV_UpperSco, Deng_Lupus, Sierra2024}. For the few disks with central depressions, the p-3-2 \htwoco~emission peaks around 150\,au for Lupus 2, 80\,au for Upper Sco 1, and 60\,au for Upper Sco 1. It is possible that for some of these disks, continuum subtraction is creating central depressions in \htwoco~by hiding \htwoco~emission in high velocity channels, thus producing a cavity in the \htwoco~emission that is not seen in the CO emission. This has been seen before in RW Aur A \citep{Kurtovic24}. Other possibilities include the presence of optically thick dust that blocks central \htwoco~emission, which is proposed to cause the central depression of \htwoco~in HD 163296 \citep{Carney2017, Hernández-Vera_2024}, or that lower energy \htwoco~lines are more sensitive to cold gas \citep{Guzman_MAPS}. Notably, Upper Sco 1 and Upper Sco 8 show central depressions in the dust radial profiles in Figure~\ref{fig:total_rp}. 

The size of the \htwoco~in R$_{90}$ is consistent between Lupus and Upper Sco, ranging from 120\,au to 230\,au, with the exception of Lupus 10, which is the largest disk in the whole AGE-PRO sample with an \htwoco~R$_{90}$ of 860\,au. 
All R$_{90}$ values are listed in Appendix Table \ref{tab:r90}. 

\begin{figure*}[!ht]
\centering
\vspace{-0.cm}
\includegraphics[width=1\textwidth]{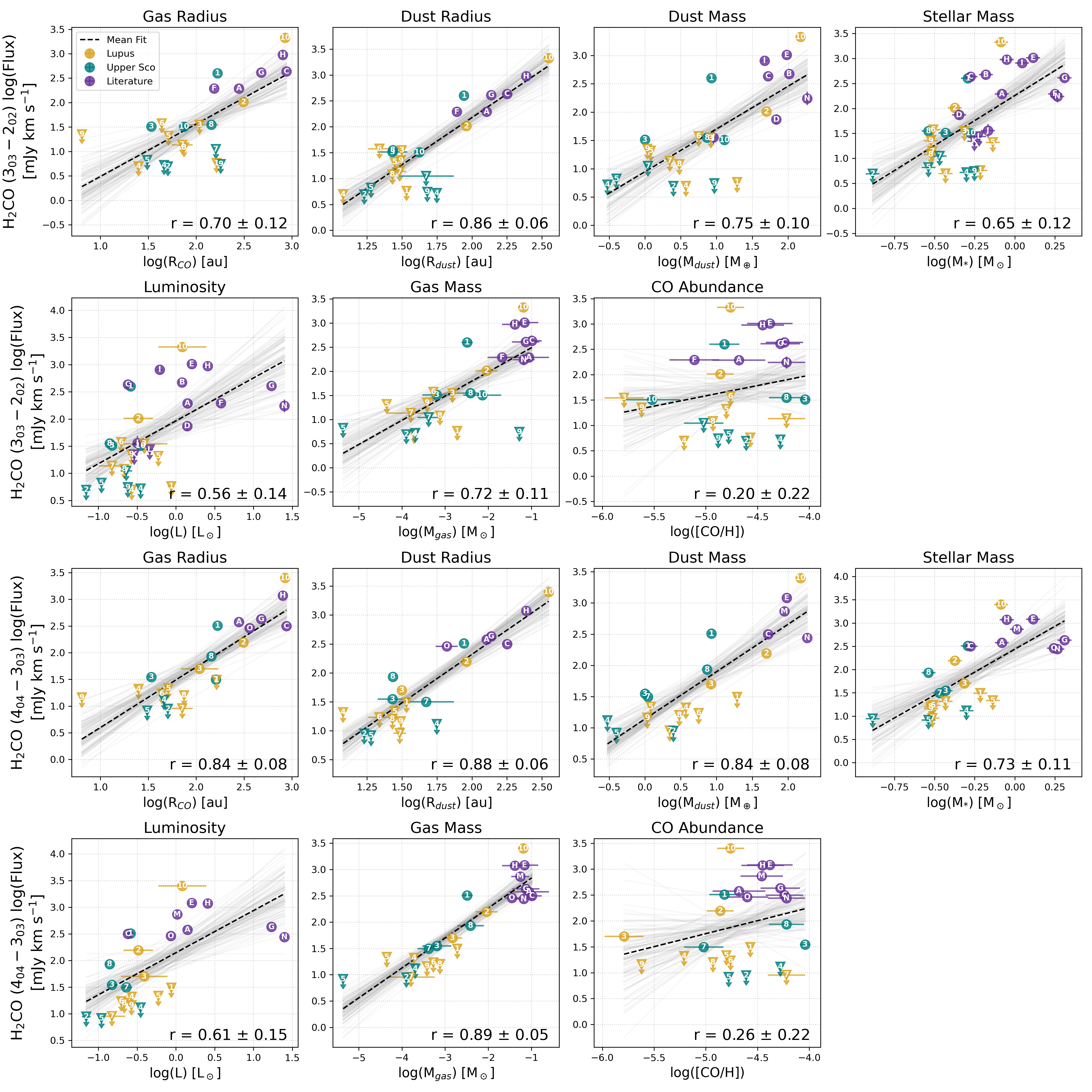}
\vspace{-0.cm}
\caption{Correlations with the \htwoco~p-3-2 and \htwoco~p-4-3 line fluxes normalized to 150 pc. All points have errorbars, but they are often too small to be visible. The correlation coefficient r from the \texttt{linmix} Python package is displayed on each plot. The gray lines show samples from the posterior of the dataset, and the dashed black line is the mean fit. Literature sources are from \cite{Pegues20}, and labels are provided in Table~\ref{tab:corr_params}. Some disks have missing data for either the line flux or disk property.}
\label{fig:corr_all}
\vspace{-0.cm}
\end{figure*}

\subsection{Excitation temperature and column density}

The excitation temperatures and total column densities are given in Table~\ref{tab:temp_and_cd}. We find excitation temperatures between 8.4-23.4\,K and column densities between 0.9-44.5$\times 10^{12}$ cm$^{-12}$ for the five disks with two \htwoco~detections. We find the highest excitation temperature of 23.4 K for Lupus 10, which is the most massive and luminous disk of the AGE-PRO sample in this work \citep{ageproXI, ageproV}. We find the highest column density of 4.45$\times 10^{13}$ cm$^{-2}$ for Upper Sco 1, which is also on the larger side in both gas and dust of the AGE-PRO disks with \htwoco~detections. The optical depths range from around 0.13 to 3.66, with Upper Sco 1 recording the highest optical depth for both the p-3-2 and p-4-3 lines. For the disks with only one \htwoco~detection, the column densities corresponding to excitation temperatures of 10 and 40 K are given in Appendix Table~\ref{tab:corr_params}. These column densities span from 0.7$\times 10^{12}$ to 6.7$\times 10^{12}$ cm$^{-12}$. From these values, we can provide estimated ranges of column densities for the disks with one \htwoco~detection.


\subsection{Correlation Analysis}

\begin{figure*}[!ht]
\centering
\vspace{-0.cm}
\includegraphics[width=0.8\textwidth]{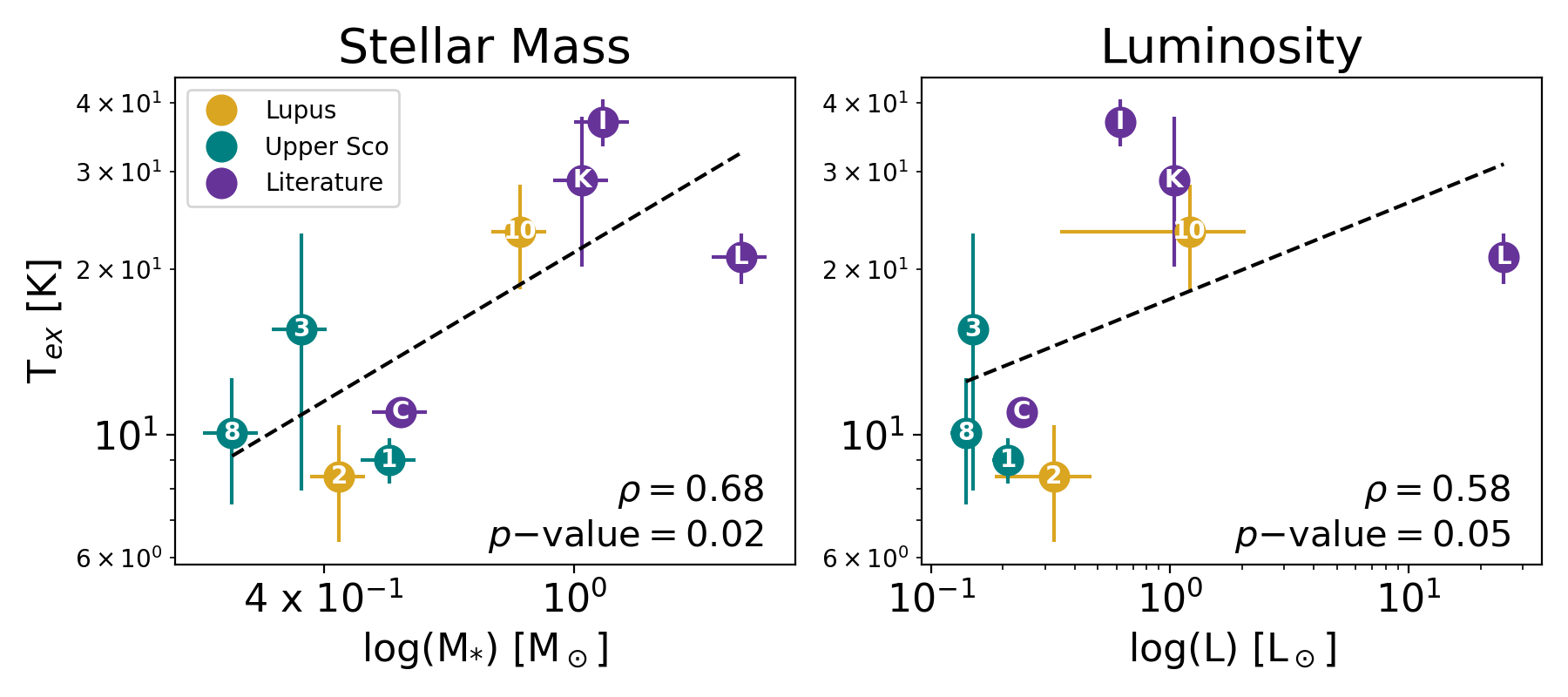}
\vspace{-0.cm}
\caption{Correlations with $T_{ex}$ and disk parameters. The excitation temperature is only found to be significantly correlated with the stellar mass and luminosity. All other correlations have a p-value greater than 0.05. While the spread of the data appears farther from the trend line compared to the line luminosity correlations, the higher errorbars likely result in a stronger correlation being possible.}
\label{fig:corr_Tex}
\vspace{-0.cm}
\end{figure*}

To investigate the origin and excitation of \htwoco~emission in disks, we examine how its line luminosity relates to the following key system properties: gas disk radius, dust disk radius, gas disk mass \citep{ageproV}, dust disk mass \citep{Deng_Lupus, AGE-PRO_IV_UpperSco}, stellar mass \citep{AGEPRO_I_overview}, stellar luminosity \citep{AGEPRO_I_overview}, and CO abundance \citep{ageproV}. The gas disk radius, $R_{\text{gas}}$, is the radius at which 90$\%$ of the CO gas emission is enclosed (taken from \cite{ageproXI}). The dust disk radius, $R_{\text{dust}}$, is the radius at which 90$\%$ of the 1.3\,mm continuum flux is enclosed (taken from \cite{ageproX}). We use the line luminosities from the 9 detections from the Lupus and Upper Sco disks in this work, as well as 13 disks from the literature \citep{Pegues20}, leading to a total sample of 22 disks. We also test correlations with these disk properties for the total column density $N_{\text{tot}}$, and the excitation temperature $T_{\text{ex}}$. For disks with only one \htwoco~detection, the two column densities corresponding to excitation temperatures of 10 and 40\,K are averaged when calculating the correlations. 

For correlations between \htwoco~line luminosity and disk properties and correlations between \htwoco~line luminosity and CO line luminosity, we use the Python package \texttt{linmix} to account for non-detections \citep{Kelly2007}. \texttt{Linmix} uses a hierarchical Bayesian model to fit a straight line to data, accounting for upper limits of non-detections. \texttt{Linmix} models the independent variable distribution as a weighted sum of K separate Gaussian distributions, and we set K=2. It produces posterior distributions for the slope, y-intercept, and intrinsic scatter by running between 5,000 and 100,000 MCMC steps. In addition to the 22 disks with significant detections, we include the \htwoco~upper limits for remaining AGE-PRO disks and from two additional disks in \cite{Pegues20} that only have upper limits for the \htwoco~p-3-2 line. All the disks used in the correlation calculations are shown in Appendix Table~\ref{tab:corr_params}. Two parameters are considered to be positively correlated if the correlation coefficient r is greater than 0.3, and very strongly positively correlated if r is greater than 0.7. If r is close to 0, then no correlation was found. The resulting correlation coefficients are given in Appendix Table~\ref{tab:corr}.

For the correlations between disk properties and excitation temperature and total column density, as well as those with the $R_{90}$ values, we only use disks with clear detections.
A Monte-Carlo Spearman-$\rho$ test with 10,000 iterations is performed to calculate correlations using the \texttt{scipy.stats.spearmanr} Python module. For each iteration, the input parameters are randomly sampled from normal distributions centered around the observed values with a standard deviation of the observed error. The Spearman rank-order correlation coefficients (denoted as $\rho$) and p-values are calculated for each iteration, and the final $\rho$ and p-values are the median values of the resulting distributions. $\rho$ is a measure of the monotonicity of the two datasets, and varies between -1 and +1. The Spearman-$\rho$ test is run with the alternative hypothesis set to ``greater", or a positive correlation. The resulting $\rho$ values and p-values are displayed in Table~\ref{tab:corr}. The disk parameters used for these correlation tests are shown in Table~\ref{tab:corr_params}. Two parameters are considered to be positively correlated if the p-value is less than 0.05.

The correlation tests between \htwoco~line luminosity and disk/stellar properties are shown in Figure~\ref{fig:corr_all}. We find that the line luminosities of both \htwoco~transitions are correlated with gas disk radius, dust disk radius, dust disk mass, gas mass, stellar luminosity, and stellar mass. Correlations for the p-4-3 line luminosity are stronger than for the p-3-2 line luminosity, with higher correlation coefficients for all correlations. We do not find correlations between \htwoco~line luminosities and the CO abundance ([CO/H]).

One possible reason why correlations are tighter for the p-4-3 line luminosity than the p-3-2 line luminosity is that the p-4-3 line is more optically thin and therefore more sensitive to the total \htwoco~gas mass. For the five disks with optical depth calculations, four out of the five disks (excluding Lupus 10, for which the optical depths are roughly equal) have a lower optical depth for the \htwoco~p-4-3 line than for the p-3-2 line. This could explain why the \htwoco~p-4-3 line luminosity has tighter correlations with disk properties. However, the \htwoco~p-4-3 line does have more detections than the \htwoco~p-3-2 line both in the AGE-PRO data and in the literature. It is possible that the tighter correlations with the p-4-3 line could just be a result of the larger sample size. 


Figure~\ref{fig:corr_all} displays the upper limits for disks with non-detections as triangles. Some of the upper limits are below the trend lines set by the clear detections. Particularly, the upper limits for Lupus 1, Upper Sco 4, Upper Sco 7, and Upper Sco 9 consistently rest below the trend line for many of the correlation plots. With the exception of Upper Sco 4, these disks do have relatively larger gas-to-dust radius ratios ($\sim$3\textendash4.8) compared to the average gas-to-dust radius ratio for the Lupus and Upper Sco disks of $\sim$2.5 \citep{AGE-PRO_IV_UpperSco, Deng_Lupus}, which could explain the lower \htwoco~emission. However, while these are the four disks that fall below the trend most significantly, other disks with upper limits also occasionally fall well below the trend line yet have lower than average gas-to-dust radius ratios. 

We find that the excitation temperature is correlated with stellar mass and stellar luminosity, and these correlations are shown in Figure~\ref{fig:corr_Tex}. No correlations are found for the total column density. For the column densities, our sample consists of the nine disks with at least one detection in the Upper Sco and Lupus regions. For the excitation temperature, only the five disks with two detections can be used. From the literature, the excitation temperature and total column density are only available for four disks. This small sample size makes it challenging to find correlations with disk parameters for the total column density and excitation temperature. It is possible that correlations with the total column density exist, but the sample is too small to significantly determine them.


\begin{figure*}[hbt!]
\centering
\vspace{-0.cm}
\includegraphics[width=1\textwidth]{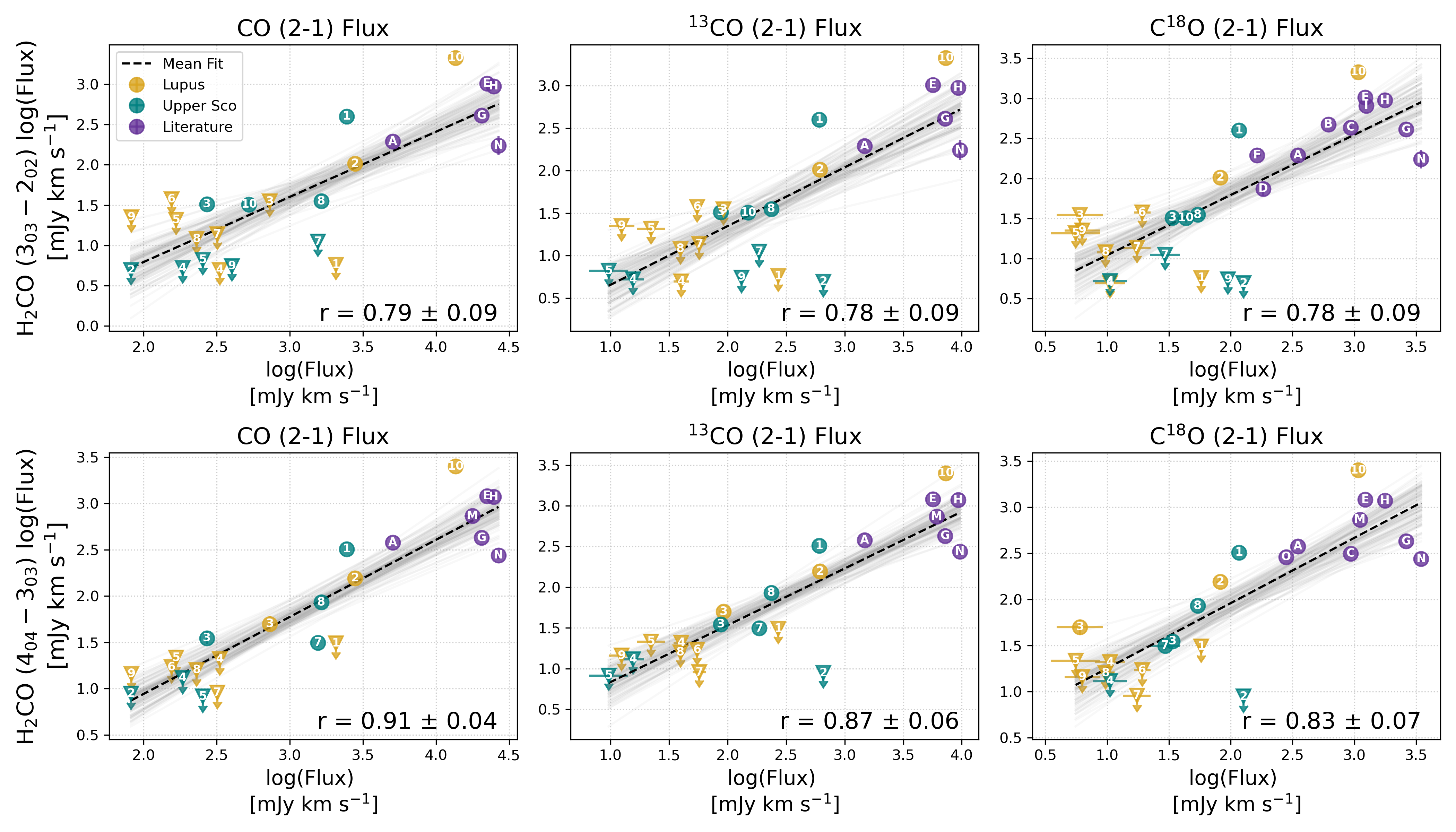}
\vspace{-0.cm}
\caption{Correlations with CO line luminosities and \htwoco~line luminosities normalized to 150 pc. Only C$^{18}$O $J$=2-1 line luminosities are recorded in \cite{Pegues20} for the disks from the literature. The CO, $^{13}$CO, and \htwoco~p-3-2 line luminosities for AS 209, GM Aur, HD 163296, IM Lup, and MWC 480 are taken from \cite{Oberg21_MAPS}.}
\label{fig:corr_CO}
\vspace{-0.cm}
\end{figure*}

We also test correlations between the \htwoco~line luminosity and the CO, $^{13}$CO, and C$^{18}$O $J$=2-1 line luminosities from \cite{Deng_Lupus, AGE-PRO_IV_UpperSco}. These results are shown in Figure~\ref{fig:corr_CO}. We find that \htwoco~line luminosity is strongly correlated with the CO, $^{13}$CO, and C$^{18}$O line luminosities. Given that CO is required to form \htwoco~through the CO ice hydrogenation mechanism on grain surfaces, we expect \htwoco~line luminosity and CO line luminosities to be correlated. As shown by Figure~\ref{fig:corr_CO}, the correlation coefficients are very similar between the CO isotopologue lines. 

\begin{figure}[htbp]
\centering
\vspace{-0.cm}
\includegraphics[width=0.45\textwidth]{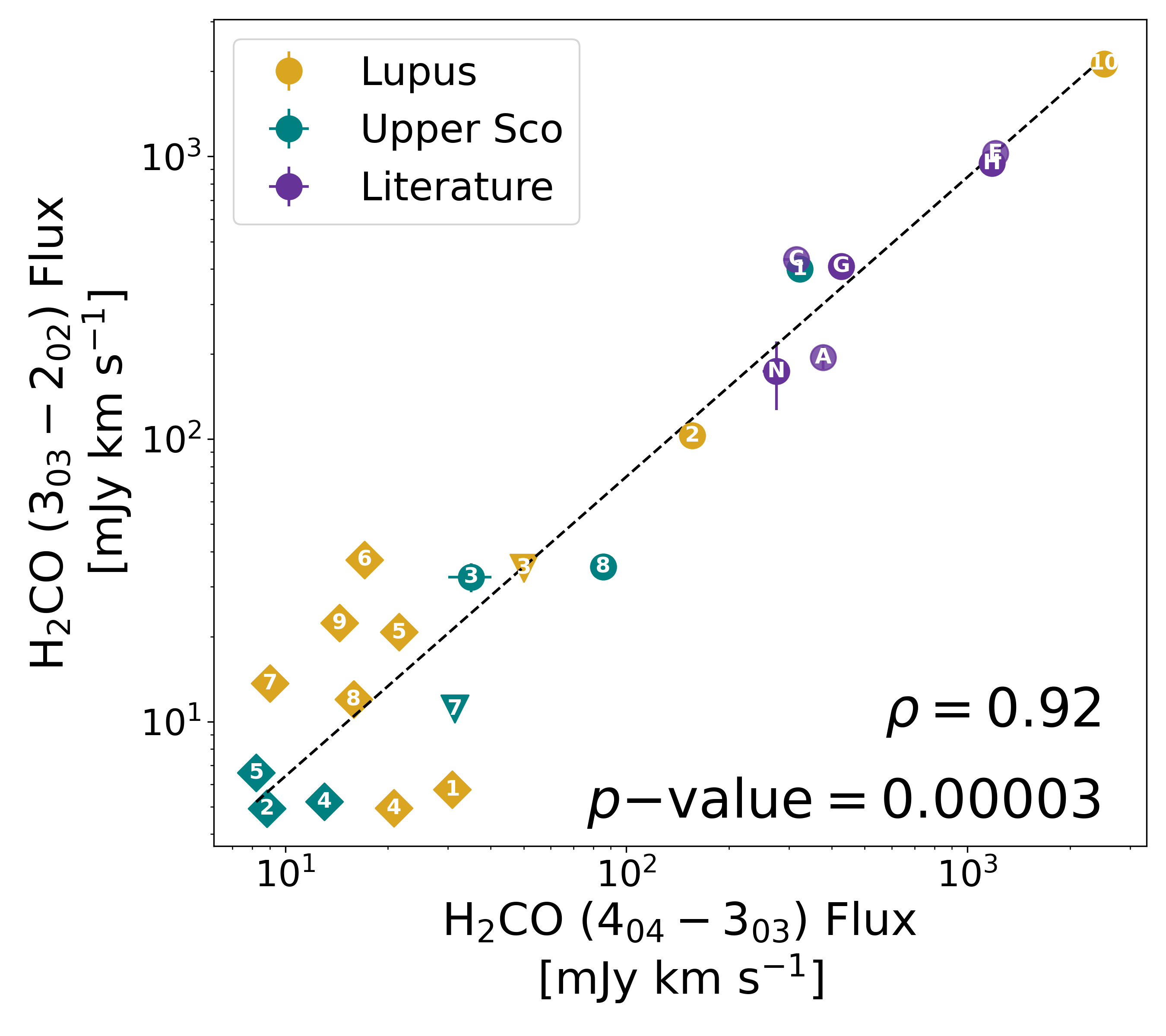}
\vspace{-0.5cm}
\caption{Correlations between \htwoco~p-3-2 line flux and \htwoco~p-4-3 line flux normalized to 150 pc. The diamonds represent the disks with 3$\sigma$ upper limit values for both the p-3-2 and p-4-3 lines, while the triangles represent the disks with detections toward p-4-3 and 3$\sigma$ upper limits toward p-3-2. We include the disks from \cite{Pegues20} with fluxes recorded for both the p-3-2 and p-4-3 lines. The dashed trend line is calculated only from the disks with significant detections in both \htwoco~lines, represented by the circle points for Lupus, Upper Sco, and the literature. The upper limit values spread farther from the trend line than the significant detections.}
\label{fig:b6vb7}
\vspace{-0.cm}
\end{figure}

We find that the \htwoco~p-3-2 line luminosity is strongly correlated with the \htwoco~p-4-3 line luminosity, as shown in Figure~\ref{fig:b6vb7}. This figure includes the 3$\sigma$ upper limit values for disks with one or no significant \htwoco~detections. 
We expect this because the p-3-2 and p-4-3 lines are from the same species, and they have similar upper energy levels and bulk temperature ranges. 

Finally, we test correlations with the p-3-2 and p-4-3 \htwoco~R$_{90}$ values. These results are shown in Appendix Figure~\ref{fig:r90_v_dust_radius}. The R$_{90}$ value is found to be borderline correlated with the \htwoco~p-4-3 line luminosity with a p-value of 0.05. For the \htwoco~p-3-2 line luminosity, the p-value is 0.11, so \htwoco~p-3-2 line luminosity and the R$_{90}$ value are not found to be correlated. However, the trend still appears to be positive and roughly linear. It is expected that the higher the \htwoco~line luminosity, the farther out into the disk the emission extends. This would result in a higher R$_{90}$ value.

\begin{table*}
\begin{tabular}{ l|lllllllll } 
 \hline
 \hline
 \htwoco~Parameter & M$_*$ & M$_{\text{gas}}$ & M$_{\text{dust}}$ & R$_{\text{gas}}$ & R$_{\text{dust}}$ & L$_{\star}$ & CO Flux & $^{13}$CO Flux & C$^{18}$O Flux \\ 
 \hline
 $L_{\rm H_2CO,3-2}$ &  \color{green}\cmark &  \color{green}\cmark & \color{green}\cmark & \color{green}\cmark & \color{green}\cmark & \color{green}\cmark & \color{green}\cmark & \color{green}\cmark & \color{green}\cmark\\
 $L_{\rm H_2CO,4-3}$ & \color{green}\cmark & \color{green}\cmark & \color{green}\cmark & \color{green}\cmark & \color{green}\cmark & \color{green}\cmark & \color{green}\cmark & \color{green}\cmark & \color{green}\cmark\\
 T$_{\text{ex}}$ & \color{green}\cmark & \color{red}\xmark & \color{red}\xmark & \color{red}\xmark & \color{red}\xmark & \color{green}\cmark & - & - & -\\
 N$_{\text{tot}}$ & \color{red}\xmark & \color{red}\xmark & \color{red}\xmark & \color{red}\xmark & \color{red}\xmark & \color{red}\xmark & - & - & - \\
 p-3-2 R$_{90}$ & - & - & - & \color{red}\xmark & \color{green}\cmark & - & - & - & -\\
 p-4-3 R$_{90}$ & - & - & - & \color{red}\xmark & \color{red}\xmark & - & - & - & -\\

\hline
\end{tabular} \\
\caption{Visualization of which \htwoco~parameters have strong correlations with each disk parameter. The green check marks represent strong correlations with p-value $<0.05$ or correlation coefficient $r>0.3$, while the red X's represent where no statistically significant correlation is found. The dashes mark where correlations were not calculated between two parameters.}
\label{tab:checkmarks}
\end{table*}

\section{Discussion}


\subsection{Possible causes of the correlation between \htwoco~luminosities and disk properties}

With the new detections of \htwoco~in 9 AGE-PRO disks, we are able to examine correlations between \htwoco~line luminosities with disk properties across a significantly expanded parameter space. As shown in Figure \ref{fig:corr_all}, the AGE-PRO sources add a population of smaller and lower-mass disks that were underrepresented in previous samples \citep[e.g.][]{Pegues20}. The broader coverage in disk sizes and masses likely contributes to our ability to identify statistically significant correlations that were not evident in prior work. A summary of the correlations we found is shown in Table \ref{tab:checkmarks}. These correlations suggest that \htwoco~emission is influenced by a combination of disk structure, mass content, and stellar irradiation. Below, we discuss several physical and chemical processes that may contribute to these trends.


First, \htwoco~line luminosities show the strongest correlations with the gas and dust disk sizes. A natural first guess might be that larger disks provide a more extended emitting area, thereby increasing the integrated line luminosity. However, we find that the \htwoco~emitting radii are only marginally correlated with the gas and dust disk sizes, which argues against the idea that emitting area alone is the dominant driver. Alternatively, the grain-surface pathway of \htwoco~formation expects \htwoco~to be abundant in the cold outer disk regions. Larger dust and gas disks may provide greater reservoirs of ices and cold CO gas which promote \htwoco~formation and survival.   

The strong correlation with dust disk mass further supports the role of dust grains in \htwoco~formation. \htwoco~is believed to form efficiently through hydrogenation of CO ice on grain surfaces. Disks with higher dust masses may host more grain surface area and larger ice reservoirs, enabling more efficient \htwoco~production. Interestingly, the \htwoco~p-4-3 line shows a strong correlation with gas disk mass, whereas the p-3-2 line shows a more moderate correlation. This could simply be a result of a difference in sample size. Additionally, the upper limit for Upper Sco 9 falls well below the trend line for the p-3-2 line, but we do not have data for the Upper Sco 9 p-4-3 line, so that outlier does not affect the correlation coefficient for gas mass and the p-4-3 line.

The observed correlation with stellar mass could arise from several contributing factors. The dust disk mass is known to scale with stellar mass \citep[e.g.,][]{,Pascucci16,Andrews18a}, and therefore the stellar mass correlation could indirectly reflect the correlation with dust disk mass. In addition, more massive stars tend to be more luminous and thus potentially increase the excitation of \htwoco. Indeed, we find a borderline correlation between stellar luminosity and excitation temperature. Although we see a strong correlation between the p-4-3 line luminosity and the stellar luminosity, the correlation with p-3-2 line luminosity is relatively weaker, which may reflect different excitation sensitivities or optical depths between the two transitions. We calculate higher optical depths for the p-3-2 line in 4/5 of the disks with detections in both lines, indicating that p-3-2 may be more likely to be optically thick.

The correlations with stellar mass and luminosity could partly reflect gas-phase \htwoco~formation in the warmer inner disk,
where higher temperatures accelerate gas-phase reactions.
However, more massive and luminous stars also tend to host larger
disks, providing more extensive cold outer regions where CO ice
hydrogenation can operate. Additionally, the AGE-PRO sources have rather low stellar luminosities (Lupus 10 is the brightest with $L=0.82L_\odot$), and the majority of the disks are $<0.5L_\odot$. As a result, we expect the mid-plane CO snowline to be inside 20 AU, which we are unable to resolve \citep{Zhang2017}. This means that the majority of the observed \htwoco~line luminosity is likely from regions outside of the mid-plane CO snowline, suggesting \htwoco~formation by CO ice hydrogenation. Even in Lupus~10, the disk with the highest
stellar mass and luminosity in our sample, the \htwoco~radial
profile shows local enhancements that coincide with dust
substructures (Figure~\ref{fig:total_rp}), suggesting that
grain-surface chemistry contributes substantially in the disk.

We find that \htwoco~line luminosities are strongly correlated with line luminosities of the $J$=2-1 transitions of CO, $^{13}$CO, and C$^{18}$O. This suggests a shared dependence on overall CO gas content. Despite the strong correlation with CO isotopologue line fluxes, we do not find a clear trend with the CO gas abundance, suggesting that \htwoco~mass is not sensitive to CO abundance. It might be because \htwoco~is only an intermediate product of CO ice processing and \htwoco~is continuously converted into more complex carbon carriers like CH$_3$OH \citep[e.g.,][]{schwarz18,Bosman18,Krijt20}. Therefore, we don't necessarily expect to see a correlation with CO abundance, so the lack of a correlation with CO abundance cannot rule out the grain-surface chemistry formation mechanism.

In summary, the correlation of \htwoco~line luminosities with dust disk radius, gas disk radius, and dust disk mass suggest that the formation of \htwoco~through grain-surface chemistry is potentially the major formation mechanism of \htwoco~in disks. This highlights the importance of large dust disks for substantial \htwoco~formation. Disks with large dust radii and masses might also be sites of more active organic chemistry, since \htwoco~will be converted to more complex organic molecules. While we can't rule out contributions from gas-phase formation, especially in the warmer inner disk, our results support the presence of a strong grain-surface component. In the next section, we will explore how the substructures in our sample also lend to this theory.

\subsection{Substructures and \htwoco~Detections}
Substructures in protoplanetary disks can lead to dust traps that allow for efficient \htwoco~formation via CO ice hydrogenation. Many of the disks in the AGE-PRO sample show evidence of substructures, which implies that dust trapping will effect the \htwoco~emission in our sample. Substructures have been detected in the azimuthally symmetric radial profiles for seven disks in our sample: Lupus 2 and 10, and Upper Sco 1, 7, 8, 9, and 10. Upper Sco 1, 7, and 8 also show inner dust cavities \citep{ageproX}. 

The set of disks with substructures compares well to the set of disks in which we detect \htwoco. In the Lupus region, we detect two \htwoco~lines toward only Lupus 2 and Lupus 10, which are the two Lupus disks with resolved substructures. We note that 7/10 Lupus disks did not have enough resolution to detect substructures, so there could be additional substructures that are undetected in the Lupus region \citep{ageproX}. In the Upper Sco region, we detect two \htwoco~lines toward Upper Sco 1, 3, and 8. Upper Sco 1 and 8 show substructures, while Upper Sco 3 does not. The remaining Upper Sco disks with substructures, Upper Sco 7, Upper Sco 9, and Upper Sco 10, have \htwoco~detected toward a single line. Overall, all seven disks in the sample with definite substructures show significant \htwoco~detections. 

The two disks with the greatest \htwoco~line luminosities (Lupus 10 and Upper Sco 1) show complex substructures that likely contribute to \htwoco~formation in dust traps. Lupus 10 and Upper Sco 1 are also both on the larger side of the sample, with Lupus 10 being the largest disk in both gas and dust and Upper Sco 1 being the third largest disk \citep{Deng_Lupus, AGE-PRO_IV_UpperSco}. Lupus 10 is an extremely structured disk, with several rings and gaps and potentially a spiral arm \citep{ageproX}. In Figure~\ref{fig:total_rp}, both the \htwoco~and the dust radial profiles show small bumps between 100\,au and 300\,au, particularly in the \htwoco~p-3-2 emission. Upper Sco 1 has a large inner dust-cavity and a ring structure, as well as a spiral arm structure \citep{ageproX, Sierra2024}. Upper Sco 1's dust ring is around 80\,au, and as shown in Figure~\ref{fig:total_rp}, this corresponds with the peak of the \htwoco~emission. We see that the dust substructures and \htwoco~are spatially linked in these disks, and we expect the substructures to exist outside the mid-plane CO snowline \citep{Zhang2017}. Through both the presence of substructures and size of the dust-disk, we see that \htwoco~formation via CO ice hydrogenation is likely to be more efficient in disks such as Lupus 10 an Upper Sco 1, resulting in the larger \htwoco~line luminosities in these disks. The parallel between substructures and \htwoco~detections supports the existence of dust traps in the disks with substructures that can accelerate \htwoco~formation on grain surfaces.

\subsection{ Potential evolution of \htwoco?}
 \begin{figure}[H]
\centering
\vspace{-0.cm}
\includegraphics[width=0.45\textwidth]{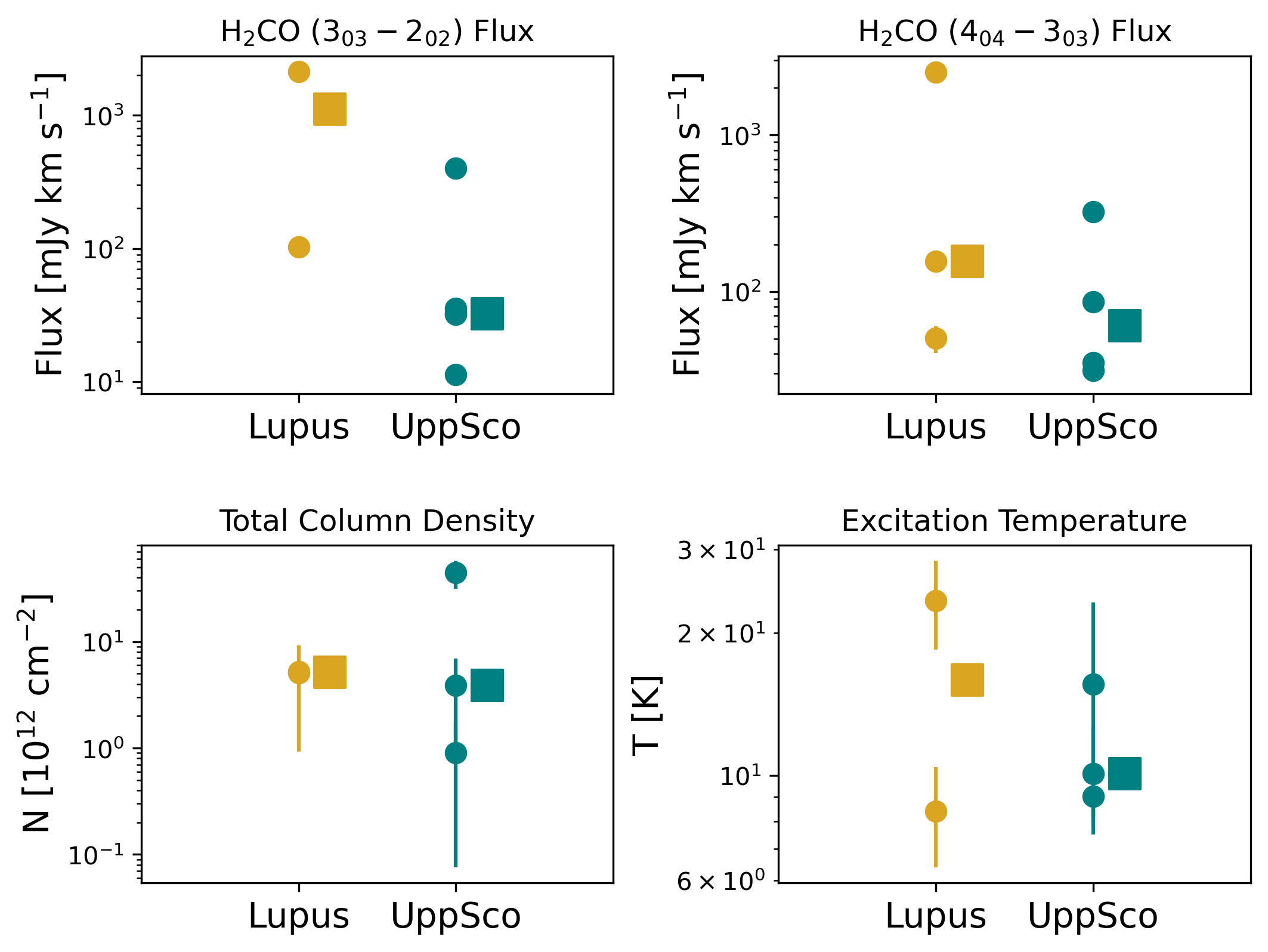}
\vspace{-0.cm}
\caption{\htwoco~line fluxes, excitation temperatures, and total column densities grouped by star-forming region. The square represents the median value for the disks in each region. We note that in the top panels, Lupus 10's large \htwoco~flux is an outlier that skews the location of the median point for the Lupus region.}
\label{fig:median_plots}
\vspace{-0.cm}
\end{figure}
In our sample, we detect \htwoco~in 3 disks in the young Lupus star forming region and in 6 disks in the relatively older Upper Sco star forming region. As shown in Figure~\ref{fig:median_plots}, the median value of the line luminosities for both \htwoco~lines decreases from the Lupus region to the Upper Sco region. However, Lupus 10 likely serves as an outlier, and we do detect \htwoco~in more relatively older Upper Sco disks than younger Lupus disks. More data points will be needed to draw a conclusion on how \htwoco~emission varies with age. Similarly, the average total column density increases and the average excitation temperature decreases from the Lupus region to the Upper Sco region. Again, too few data points are available to confirm these relationships. With only five disks with significant \htwoco~detections in both lines, we cannot come to a clear conclusion on how \htwoco~evolves with disk age.


\section{Conclusion}
In this paper, we present an analysis of \htwoco~toward 20 Class II disks in the Lupus and Upper Scorpius star-forming regions and measure line luminosities of two \htwoco~lines. Both \htwoco~lines are robustly detected in five disks, and one \htwoco~line is detected in four additional disks.  The rotational diagram method is used to calculate the excitation temperature and total column density for the disks with two \htwoco~line detections, and these values are corrected for optical depth. We use the \texttt{linmix} Python package and a Spearman-$\rho$ correlation test to search for correlations between the line luminosity, excitation temperature, and total column density and various disk parameters. The following summarizes our conclusions:
\begin{itemize}

    \item We find significant \htwoco~detections in 9 of the 20 AGE-PRO disks in our sample. The p-3-2 \htwoco~line is detected at a rate of 20\% in the Lupus region and 50\% in the Upper Sco region. The p-4-3 \htwoco~line is detected at a rate of 30\% in the Lupus region and 40\% in the Upper Sco region.
    
    \item We calculate excitation temperatures and total column densities for five disks, finding values between 8.4-23.4 K and 0.9-44.5$\times$10$^{12}$ cm$^{-2}$, respectively. 
    
    \item We find the R$_{90}$ of \htwoco~line emission between 121-830\,au for the Lupus region and between 134-229\,au for the Upper Sco sample.
    
    \item We find that the \htwoco~line luminosity is strongly correlated with gas disk radius, dust disk radius, dust disk mass, stellar luminosity, stellar mass, and gas mass for both the \htwoco~p-3-2 and the \htwoco~p-4-3 transitions. 
     
    \item The \htwoco~line luminosity has the strongest correlation with the gas and dust disk radii, suggesting that larger gas and dust disks have larger reservoirs of ices and CO gas where \htwoco~can form. The strong correlation with dust disk mass suggests that the presence of more dust in the outer regions of the disk could allow for more efficient surface-grain chemistry as a formation pathway for \htwoco.

    \item The \htwoco~line luminosities are strongly correlated with the line fluxes of J=2-1 CO, $^{13}$CO, and C$^{18}$O lines, but not with CO abundance ([n$_{\rm CO}$/n$_{\rm H}$]). This suggests that \htwoco~emission depends on the total amount of CO gas but is not overly sensitive to the CO abundance.
    
    \item The excitation temperature of \htwoco~is found to be strongly correlated with the stellar mass and luminosity. More massive and luminous stars potentially increase \htwoco~excitation.

    \item The disks with substructures are more likely to have significant \htwoco~detections, supporting that dust traps in these substructures can lead to more \htwoco~formation via CO ice hydrogenation.
    
    \item More data is needed to determine the relationship between \htwoco~emission and disk age. Improving the precision of our age estimates for protoplanetary disks would allow more robust analysis of the evolution of 
    \htwoco~over time. 
    
\end{itemize}
Overall, we find that larger protoplanetary disks with more massive and luminous stars will likely have more \htwoco~emission. The disks with the largest \htwoco~line luminosities also have clear substructures, suggesting that dust traps may form in these disks and accelerate \htwoco~formation on dust grains. Future work could attempt to detect more complex organic molecules to determine if these detections follow the same trends that we find for \htwoco. Additionally, while there are not enough disks with significant \htwoco~detections in our sample to determine the relationship between \htwoco~emission and disk age, future research could add to this sample by continuing to study \htwoco~systematically in disks that span the lifetime of protoplanetary disks.

\begin{acknowledgments}
The authors are grateful to the anonymous reviewer for their constructive feedback, which helped enhance this work. This paper makes use of the following ALMA data: ADS/JAO.ALMA\#2021.1.00128.L. ALMA is a partnership of ESO (representing its member states), NSF (USA) and NINS (Japan), together with NRC (Canada), MOST and ASIAA (Taiwan), and KASI (Republic of Korea), in cooperation with the Republic of Chile. The Joint ALMA Observatory is operated by ESO, AUI/NRAO and NAOJ. The National Radio Astronomy Observatory is a facility of the National Science Foundation operated under cooperative agreement by Associated Universities, Inc.

E.C. and K.Z. acknowledge the support of the NSF AAG grant \#2205617. E.C. and K.Z. also acknowledge the Sophomore Research Fellowship from the University of Wisconsin-Madison, which was also used to support this research. N.T.K acknowledges the support of the Deutsche Forschungsgemeinschaft (DFG, German Research Foundation) - 325594231, FOR 2634/2. P.P and A.S acknowledge the support from the UK Research and Innovation (UKRI) under the UK government's Horizon Europe funding guarantee from ERC (under grant agreement No 101076489). A.S. also acknowledges support from FONDECYT de Postdoctorado 2022 \#3220495. D.D. acknowledges support from Collaborative NSF Astronomy \& Astrophysics Research grant (ID: 2205870). C.A.G. acknowledges support from FONDECYT de Postdoctorado 2021 \#3210520. N.T.K. and P.P. acknowledge the support from the Alexander von Humboldt Foundation in the framework of the Sofja Kovalevskaja Award endowed by the Federal Ministry of Education and Research. J.M. acknowledges support from ANID -- Millennium Science Initiative Program -- Center Code NCN2024\_001.

\end{acknowledgments}
\newpage

\appendix
\restartappendixnumbering

\section{Additional Tables}
Table \ref{tab:mask_params} shows the Keplerian mask parameters for all Lupus and Upper Sco sources (except for Upper Sco 6, which does not have \htwoco~data at the time of this paper). The Keplerian mask parameters for the Lupus sources are from \citep{Deng_Lupus}. The incl, PA, and distance for the Upper Sco sources were obtained from \cite{Deng_Lupus} for the Lupus sources and from \cite{AGE-PRO_IV_UpperSco} for the Upper Sco sources, with some empirical adjustments made based on the CO emission. The v$_{\text{sys}}$ values for Upper Sco were found using the \texttt{eddy} Python package \citep{eddy}. The remaining parameters were empirically determined by applying the mask to the CO emission. The mask radius was occasionally decreased for the \htwoco~emission for cases in which the detection was clear and the mask appeared too large. These parameters do not represent the true physical properties of the disks.

\vspace{0.4cm}
\begin{table}[h] 
\centering
\hspace{-1.8cm}\begin{tabular}{ lllllll } 
 \hline
 \hline
 Disk & PA & incl & M$_{\star}$ & Dist & v$_{\text{sys}}$ & CO Mask \\ 
 & & & & & & Radius\\
  & ($^\circ$) & ($^\circ$) & (M$_\odot$) & (pc) & (km s$^{-1}$) & (")  \\ 
 \hline
 Lupus 1 & 290 & 62 & 0.84 & 153.5 & 4.4 & 1.6\\ 
 Lupus 2 & 35 & 35 & 0.46 & 155.2 & 3.6 & 2.5\\ 
 Lupus 3 & 195 & 57 & 0.52 & 159.8 & 4.6 & 1.5  \\
 Lupus 4 & 225 & 30 & 0.42 & 156.7 & 3.6 & 0.5\\ 
 Lupus 5 & 43 & 45 & 0.87 & 155.3 & 3.8 & 0.5\\ 
 Lupus 6 & 284 & 53 & 0.32 & 163.0 & 4.2 & 0.5\\ 
 Lupus 7 & 342 & 43 & 0.33 & 160.6 & 3.4 & 0.7 \\ 
 Lupus 8 & 270 & 45 & 0.32 & 155.9 & 4.3 & 0.7\\ 
 Lupus 9 & 120 & 80 & 1.0 & 160.5 & 3.1 & 1.2\\ 
 Lupus 10 & 108 & 53 & 1.21 & 158.0 & 5.4 & 8.0\\ 
 UpperSco 1 & 43 & 37 & 0.5 & 132.0 & 4.5 & 1.5  \\ 
 UpperSco 2 & 63 & 51 & 1.3 & 138.7 & 3.4 & 1.25  \\ 
 UpperSco 3 & 80 & 57 & 1.5 & 140.7 & 3.5 & 1.0\\
 UpperSco 4 & 265 & 60 & 1.3 & 137.0 & 4.5 & 0.45  \\ 
 UpperSco 5 & 71 & 56 & 0.55 & 144.0 & 5.1 & 1.0  \\ 
 UpperSco 7 & 188 & 27 & 0.65 & 153.0 & 4.2 & 2.0\\ 
 UpperSco 8 & 18 & 60 & 0.3 & 139.9 & 3.3 & 2.0\\
 UpperSco 9 & 303 & 71 & 0.65 & 137.0 & 5.1 & 1.0\\ 
 UpperSco 10 & 331 & 54 & 1.3 & 136.9 & 4.3 & 1.0\\  
\hline
\end{tabular} \\
\caption{The Keplerian mask parameters for the Lupus and Upper Sco sources.}
\label{tab:mask_params}
\end{table}

\newpage
Table \ref{tab:r90} shows the R$_{90}$ values for the Lupus and Upper Sco sources with significant \htwoco~detections. The R$_{90}$ values were calculated by finding the radius at which the enclosed flux reached 90$\%$ of the total cumulative flux.

\begin{table}[h]
\centering
\hspace{-1.8cm}\begin{tabular}{ lllllll } 
 \hline
 \hline
 Disk & H$_2$CO Line & R90 Value & R90 Value \\ 
  &  & (") & (AU) \\ 
 \hline
 Lupus 2 &  p-3-2 & 1.65$_{-0.10}^{+0.20}$ & 256.1$_{-15.5}^{+31.0}$ \\ 
 ... &  p-4-3 & 2.25$_{-0.30}^{+0.40}$ & 349.2$_{-46.6}^{+62.1}$  \\ 
 Lupus 3 & p-4-3 & 1.55$_{-0.60}^{+0.70}$ & 247.7$_{-95.9}^{+111.9}$  \\ 
 Lupus 10 & p-3-2 & 5.25$_{-0.10}^{+0.10}$ & 829.5$_{-15.8}^{+15.8}$ \\ 
 ... &  p-4-3 & 4.35$_{-0.10}^{+0.00}$ & 687.3$_{-15.8}^{+0.0}$ \\ 
 UpperSco 1 & p-3-2 & 1.65$_{-0.00}^{+0.10}$ & 217.8$_{-0.0}^{+13.2}$  \\ 
  ... &  p-4-3 & 1.65$_{-0.00}^{+0.10}$ & 217.8$_{-0.0}^{+13.2}$ \\
 UpperSco 3 & p-3-2 & 1.55$_{-0.30}^{+0.30}$ & 218.1$_{-42.2}^{+42.2}$\\ 
 ... &  p-4-3 & 1.35$_{-0.30}^{+0.60}$ & 189.9$_{-42.2}^{+84.4}$\\ 
 UpperSco 7 & p-4-3 & 0.95$_{-0.20}^{+0.60}$ & 145.4$_{-30.6}^{+91.8}$\\ 
 UpperSco 8 & p-3-2 & 1.05$_{-0.00}^{+0.10}$ & 146.0$_{-0.0}^{+13.9}$\\ 
 ... &  p-4-3 & 1.65$_{-0.20}^{+0.20}$ & 229.4$_{-27.8}^{+27.8}$ \\ 
 UpperSco 9 & p-3-2 & 1.55$_{-0.10}^{+0.20}$ & 212.4$_{-13.7}^{+27.4}$\\
 UpperSco 10 & p-3-2 & 0.85$_{-0.20}^{+0.40}$ & 134.3$_{-31.6}^{+63.2}$\\  
\hline
\end{tabular} \\
\caption{The R$_{90}$ values for the Lupus and Upper Sco sources.}
\label{tab:r90}
\end{table}

\newpage
Table \ref{tab:corr} shows the $\rho$ and p-values determined using a Spearman-$\rho$ statistical test and the correlation coefficients r from the \texttt{linmix} correlation calculations that include the upper limits for non-detections. The significant p-values ($<$0.05) and correlation coefficients ($>$0.3) are bolded, signifying that a statistically probable correlation was found. The statistical test and its results are discussed in \S 3.3.

\begin{table}[h]
\centering
\hspace{-1.95cm}\begin{tabular}{ lllll } 
 \hline
 \hline
 Disk & H$_2$CO & $\rho$ & p-value & r \\ 
 Parameter & Parameter & & &\\
 \hline
 M$_*$ & $L_{\rm H_2CO,3-2}$ 
 &  &  & \textbf{0.65 $\pm$ 0.12}\\ 
 ... &  $L_{\rm H_2CO,4-3}$ &  &   & \textbf{0.73 $\pm$ 0.11}\\ 
 ... &  T$_{ex}$ & 0.68$_{-0.18}^{+0.10}$ & \textbf{0.02}$_{\textbf{-0.01}}^{\textbf{+0.06}}$  \\ 
 ... &  N$_{\text{tot}}$ & 0.13$_{-0.19}^{+0.19}$ & 0.33$_{-0.19}^{+0.24}$  \\ 
 \hline
 M$_{\text{gas}}$ &  $L_{\rm H_2CO,3-2}$ &  &  & \textbf{0.72 $\pm$ 0.11}\\ 
 ... &  $L_{\rm H_2CO,4-3}$ &  &  & \textbf{0.89 $\pm$ 0.05} \\ 
 ... &  T$_{\text{ex}}$ & 0.10$_{-0.20}^{+0.50}$ & 0.44$_{-0.29}^{+0.13}$  \\ 
 ... &  N$_{\text{tot}}$ & 0.15$_{-0.17}^{+0.13}$ & 0.35$_{-0.12}^{+0.17}$  \\ 
 \hline
 M$_{\text{dust}}$ &  $L_{\rm H_2CO,3-2}$ &  &  & \textbf{0.75 $\pm$ 0.10}\\ 
 ... &  $L_{\rm H_2CO,4-3}$ &  &  & \textbf{0.84 $\pm$ 0.08}\\ 
 ... &  T$_{\text{ex}}$ & 0.43$_{-0.23}^{+0.18}$ & 0.12$_{-0.08}^{+0.18}$  \\ 
 ... &  N$_{\text{tot}}$ & 0.19$_{-0.19}^{+0.16}$ & 0.26$_{-0.15}^{+0.24}$  \\ 
 \hline
 R$_{\text{gas}}$ &  $L_{\rm H_2CO,3-2}$ &  &  & \textbf{0.70 $\pm$ 0.12}\\ 
 ... &  $L_{\rm H_2CO,4-3}$ &  &  & \textbf{0.84 $\pm$ 0.08}\\ 
 ... &  T$_{\text{ex}}$ & 0.14$_{-0.29}^{+0.34}$ & 0.39$_{-0.23}^{+0.21}$  \\ 
 ... &  N$_{\text{tot}}$ & 0.16$_{-0.25}^{+0.17}$ & 0.33$_{-0.15}^{+0.27}$  \\ 
 ... &  p-3-2 R$_{90}$ & 0.68$_{-0.21}^{+0.21}$ & \textbf{0.05}$_{\textbf{-0.06}}^{\textbf{+0.12}}$  \\ 
 ... &  p-4-3 R$_{90}$ & 0.61$_{-0.21}^{+0.18}$ & 0.07$_{-0.06}^{+0.12}$  \\ 
 \hline
 R$_{\text{dust}}$ &  $L_{\rm H_2CO,3-2}$ & &  & \textbf{0.86 $\pm$ 0.06}\\ 
 ... &  $L_{\rm H_2CO,4-3}$ &  &  & \textbf{0.88 $\pm$ 0.06}\\ 
 ... &  T$_{\text{ex}}$ & 0.26$_{-0.23}^{+0.34}$ & 0.31$_{-0.21}^{+0.17}$  \\ 
 ... &  N$_{\text{tot}}$ & 0.32$_{-0.22}^{+0.22}$ & 0.18$_{-0.13}^{+0.21}$  \\
 ... &  p-3-2 R$_{90}$ & 0.75$_{-0.21}^{+0.14}$ & \textbf{0.03}$_{\textbf{-0.02}}^{\textbf{+0.08}}$  \\ 
 ... &  p-4-3 R$_{90}$ & 0.57$_{-0.21}^{+0.18}$ & 0.09$_{-0.06}^{+0.13}$  \\ 
 \hline
 L$_*$ &  $L_{\rm H_2CO,3-2}$ &  &  & \textbf{0.56 $\pm$ 0.14}\\ 
 ... &  $L_{\rm H_2CO,4-3}$ &  &  & \textbf{0.61 $\pm$ 0.15}\\ 
 ... &  T$_{\text{ex}}$ & 0.58$_{-0.22}^{+0.13}$ & \textbf{0.05}$_{\textbf{-0.03}}^{\textbf{+0.12}}$  \\ 
 ... &  N$_{\text{tot}}$ & 0.10$_{-0.21}^{+0.21}$ & 0.37$_{-0.22}^{+0.27}$  \\ 
\hline

[CO/H] &  $L_{\rm H_2CO,3-2}$ &  &  & 0.20 $\pm$ 0.22\\ 
 ... &  $L_{\rm H_2CO,4-3}$ & &  & 0.26 $\pm$ 0.22\\  
 ... &  T$_{\text{ex}}$ & 0.20$_{-0.50}^{+0.40}$ & 0.37$_{-0.23}^{+0.31}$  \\
 ... &  N$_{\text{tot}}$ & -0.03$_{-0.25}^{+0.25}$ & 0.53$_{-0.25}^{+0.24}$  \\ 
\hline

CO Flux &  $L_{\rm H_2CO,3-2}$ &  &  & \textbf{0.79 $\pm$ 0.09}\\ 
 ... &  $L_{\rm H_2CO,4-3}$ &  &  & \textbf{0.91 $\pm$ 0.04}\\ 
\hline
$^{13}$CO Flux &  $L_{\rm H_2CO,3-2}$ & &  & \textbf{0.78 $\pm$ 0.09}\\ 
 ... &  $L_{\rm H_2CO,4-3}$ & &  & \textbf{0.87 $\pm$ 0.06}\\ 
\hline
C$^{18}$O Flux &  $L_{\rm H_2CO,3-2}$ & & & \textbf{0.78 $\pm$ 0.09}\\ 
 ... &  $L_{\rm H_2CO,4-3}$ &  &  & \textbf{0.83 $\pm$ 0.07}\\  
\hline
\end{tabular} \\
\caption{The results from the Spearman-$\rho$ statistical test and the \texttt{linmix} analysis to determine correlations between \htwoco~line flux, excitation temperature, and total column density and disk parameters.}
\label{tab:corr}
\end{table}

\newpage
Table \ref{tab:corr_params} displays all parameters used to calculate the correlations. The first column shows the letters that represent each disk from the literature in the correlation plots. The R$_{\text{CO}}$ values for the disks from the literature are from \cite{Long22_size}, and the gas masses and CO abundances are from \cite{Trapman2025}. The CO, C$^{18}$O, and $^{13}$CO line fluxes for IM Lup, GM Aur, AS 209, HD 163296, and MWC 480 are from \cite{Oberg21_MAPS}. 

\begin{table*}[h!]
\centering
\scalebox{0.65}{
\hspace{-9cm}\begin{tabular}{llllllllllllllllllllllllllllllllllllllll}
\hline
\hline
& & \multicolumn{2}{c}{\htwoco~Fluxes} & \multicolumn{7}{c}{Stellar Properties}  & & & & & \multicolumn{3}{c}{CO Fluxes} & \\
\hline
&Disk & p-3-2 & p-4-3 & M$_{\star}$ & L$_{\star}$ & R$_{\text{CO}}$ & R$_{\text{dust}}$ & M$_{\text{dust}}$ & log(M$_{\text{gas}}$) & log($\chi_{CO}$) & T$_{\text{ex}}$ & N & R$_{90}$ p-3-2 & R$_{90}$ p-4-3 & CO 2-1 & C$^{18}$O 2-1 & $^{13}$CO 2-1 & Refs. \\
& & (mJy km  & (mJy km  & (M$_\odot$) & (L$_\odot$) &(au)&(au)&(M$_{\oplus}$) & (M$_{\odot}$) & & (K) & ($10^{12}$ cm$^{-2}$) & (au) & (au) & (mJy km &(mJy km &(mJy km & \\
& & s$^{-1}$) & s$^{-1}$) & &  & & & & &  &  & &  &  & s$^{-1}$)& s$^{-1}$)& s$^{-1}$) & \\
\hline
        A&AS 209 & 299 & 580 & 0.83 & 1.41 & 280 & 127 & - & -1.06 & -4.68 & - & - & - & - & 7790 & 538 & 2269 & 1, 2, 3, 9\\ 
        B&CI Tau & 420 & - & 0.66 & 1.2 & - & - & 103.42 & - & - & - & - & - & - & - & 549 & - & 1\\ 
        C&DM Tau & 462 & 337 & 0.53 & 0.24 & 876 & 178 & 52.42 & -0.99 & -4.24 & 11 & 2.4 & - & - & - & 998 & - & 1, 3, 9\\ 
        D&DO Tau & 85 & - & 0.45 & 1.4 & - & - & 68.54 & - & - & - & - & - & - & - & 210 & - & 1\\ 
        E&GM Aur & 913 & 1074 & 1.3 & 1.6 & - & - & 95.92 & -1.16 & -4.38 & - & - & - & - & 19844 & 1092 & 5028 & 1, 2, 9\\ 
        F&HD 143006 & 161 & - & 1.78 & 3.8 & 154 & 78 & - & -1.68 & -4.38 & - & - & - & - & - & 135 & - & 1, 3, 9\\ 
        G&HD 163296 & 902 & 942 & 2.04 & 17 & 478 & 137 & - & -1.68 & -5.11 & - & - & - & - & 45246 & 5783 & 15885 & 1, 2, 3, 9\\ 
        H&IM Lup & 852 & 1063 & 0.89 & 2.57 & 803 & 244 & - & -1.39 & -4.45 & - & - & - & - & 22342 & 1592 & 8370 & 1, 2, 3, 9\\ 
        I&J1604-2130 & 821 & - & 1.11 & 0.62 & - & - & 46.9 & - & - & 37 & 21 & - & - & - & 1267 & - & 1\\ 
        J&J1609-1908 & 43 & - & 0.68 & 0.32 & - & - & 9.06 & - & - & - & - & - & - & - & - & - & 1\\ 
        K&LkCa 15 & - & 662 & 1.03 & 1.04 & - & - & 88.67 & -1.26 & -4.46 & 29 & 3.1 & - & - & 16000 & 1000 & 5500 & 1, 4, 9\\ 
        L&MWC 480 & 151 & 239 & 1.84 & 25 & - & - & 184.65 & -1.18 & -4.22 & 21 & 1.9 & - & - & 23226 & 3017 & 8361 & 1, 2, 9\\ 
        M&V4046 Sgr & - & 1218 & 1.75 & 0.86 & 362 & 66 & - & -1.46 & -4.60 & - & - & - & - & - & 1184 & - & 1, 3, 9\\ 

        N&J1612-1859 & $<$26 & - & 0.56 & 0.29 & - & - & - & - & - & - & - & - & - & - &  & - & 1\\ 

        O&J1614-1906 & $<$30 & - & 0.6 & 0.46 & - & - & - & - & - & - & - & - & - & - & - & - & 1\\ 
        
        &Lupus 1 & $<$5.49 & $<$29.49 & 0.61 & 0.87 & 163 & 34.1 & 19.4 & -2.72 & -4.57 & - & - & - & - & 1979.9 & 55.3 & 259.1 & 5, 6, 8\\ 
        
        &Lupus 2 & 96.2 & 145.8 & 0.42 & 0.33 & 313 & 91.1 & 50.0 & -2.04 & -4.86 & 8.40 & 5.06 & 256.1 & 349.2 & 2602.3 & 77 & 574.4 & 5, 6, 8\\ 
        &Lupus 3 & $<$30.81 & 44.1 & 0.49 & 0.39 & 109 & 31.5 & 8.4 & -2.83 & -5.79 & [10, 40] & [1.60, 0.80] & - & 247.7 & 643.8 & 5.3 & 81.5 & 5, 6, 8\\ 

        &Lupus 4 & $<$4.53 & $<$19.05 & 0.37 & 0.27 & 25.1 & 12 & 3.7 & -3.72 & -5.21 & - & - & - & - & 305.1 & 9.7 & 36.9 & 5, 6, 8\\ 

        &Lupus 5 & $<$19.38 & $<$20.1 & 0.73 & 0.59 & 51.4 & 5.1 & 1.2 & -4.35 & -4.80 & - & - & - & - & 155.7 & 5.2 & 20.8 & 5, 6, 8\\ 

        &Lupus 6 & $<$31.68 & - & 0.31 & 0.20 & 43.9 & 21.8 & 5.7 & -3.27 & -4.76 & - & - & - & - & 131.7 & 16.3 & 46.7 & 5, 6, 8\\ 

        &Lupus 7 & $<$11.85 & $<$7.86 & 0.31 & 0.15 & 71.5 & 30.4 & 2.2 & -3.79 & - 4.22& - & - & - & - & 278.6 & 15.2 & 50.1 & 5, 6, 8\\ 

        &Lupus 8 & $<$11.1 & $<$14.7 & 0.30 & 0.22 & 74.7 & 27 & 3.0 & -3.12 & -4.93 & - & - & - & - & 213.9 & 9.0 & 36.7 & 5, 6, 8\\ 

        &Lupus 9 & $<$19.53 & $<$12.6 & 0.30 & 0.27 & 6.42 & 31 & 1.1 & -3.42 & -5.62 & - & - & - & - & 72.2 & 5.5 & 10.9 & 5, 6, 8\\ 
        
        &Lupus 10 & 1916.7 & 2269.4 & 0.82 & 1.21 & 850 & 353 & 149.3 & -1.19 & -4.76 & 23.40 & 5.18 & 829.5 & 687.3 & 12273.4 & 970 & 6574 & 5, 6, 8\\ 
        &Upper Sco 1 & 517.46 & 416.83 & 0.51 & 0.25 & 167 & 87.6 & 8.5 & -2.49 & 9.02 & 44.54 & 217.8 & 217.8 & 3171.6 & 151.3 & 781.8 & 5, 7, 8\\ 

        &Upper Sco 2 & $<$5.85 & $<$10.5 & 0.13 & 0.07 & 50.6 & 17 & 2.5 & -3.9 & -4.61 & - & - & - & - & 97.6 & 151.3 & 781.8 & 5, 7, 8\\ 
        
        &Upper Sco 3 & 37.53 & 40.5 & 0.37 & 0.15 & 34 & 27 & 1 & -3.19 & -4.04 & 15.56 & 0.91 & 218.1 & 190 & 312.8 & 39 & 101.3 & 5, 7, 8\\ 

        &Upper Sco 4 & $<$6.27 & $<$15.63 & 0.50 & 0.35 & 46.3 & 56 & 0.3 & -3.69 & -4.28 & - & - & - & - & 222.0 & 12.7 & 18.7 & 5, 7, 8\\ 

        &Upper Sco 5 & $<$7.17 & $<$8.91 & 0.29 & 0.11 & 30.8 & 19 & 0.4 & -4.78 & -4.78 & - & - & - & - & 275.4 & - & 10.5 & 5, 7, 8\\ 
         
        &Upper Sco 7 & $<$10.7 & 30.27 & 0.34 & 0.23 & 160 & 47 & 1.1 & -3.38 & -5.02 & [10, 40] & [3.47, 1.48] & - & 145.4 & 1506.3 & 28.4 & 180.3 & 5, 7, 8\\
        &Upper Sco 8 & 41.17 & 99.63 & 0.29 & 0.14 & 144 & 27.2 & 7.4 & -2.41 & -4.22 & 10.08 & 3.87 & 146 & 229.4 & 1917 & 63 & 275.4 & 5, 7, 8\\ 
        &Upper Sco 9 & 13.62 & - & 0.56 & 0.24 & 139 & 48 & 9.3 & -1.28 & -4.88 & [10, 40] & [0.32, 0.50] & 212.4 & - & 481.3 & 114.2 & 157.5 & 5, 7, 8\\
        &Upper Sco 10 & 38.55 & - & 0.53 & 0.35 & 75 & 42 & 12.9 & -2.14 & -5.52 & [10, 40] & [6.74, 5.40] & 134.3 & - & 630.8 & 52 & 180 & 5, 7, 8\\ 
        \hline
\end{tabular}
}
\caption{The parameters used to calculate correlations for both the AGE-PRO and literature sources. Uncertainties can be found in the original papers, which are cited below.}
\label{tab:corr_params}
\tablerefs{The Refs. column gives references for each disk:   1 = \citet{Pegues20}, 2 = \citet{Oberg21_MAPS}, 3 = \citet{Long22_size}, 4 = \citet{LkCa23}, 5 = \citet{ageproX}, 6 = \citet{Deng_Lupus}, 7 = \citet{AGE-PRO_IV_UpperSco}, 8 = \citet{ageproV}, 9 = \citet{Trapman2025}}
\end{table*}
\newpage

\section{Additional Figures}
\begin{figure*}[hb]
\centering
\vspace{-0.cm}
\includegraphics[width=0.9\textwidth]{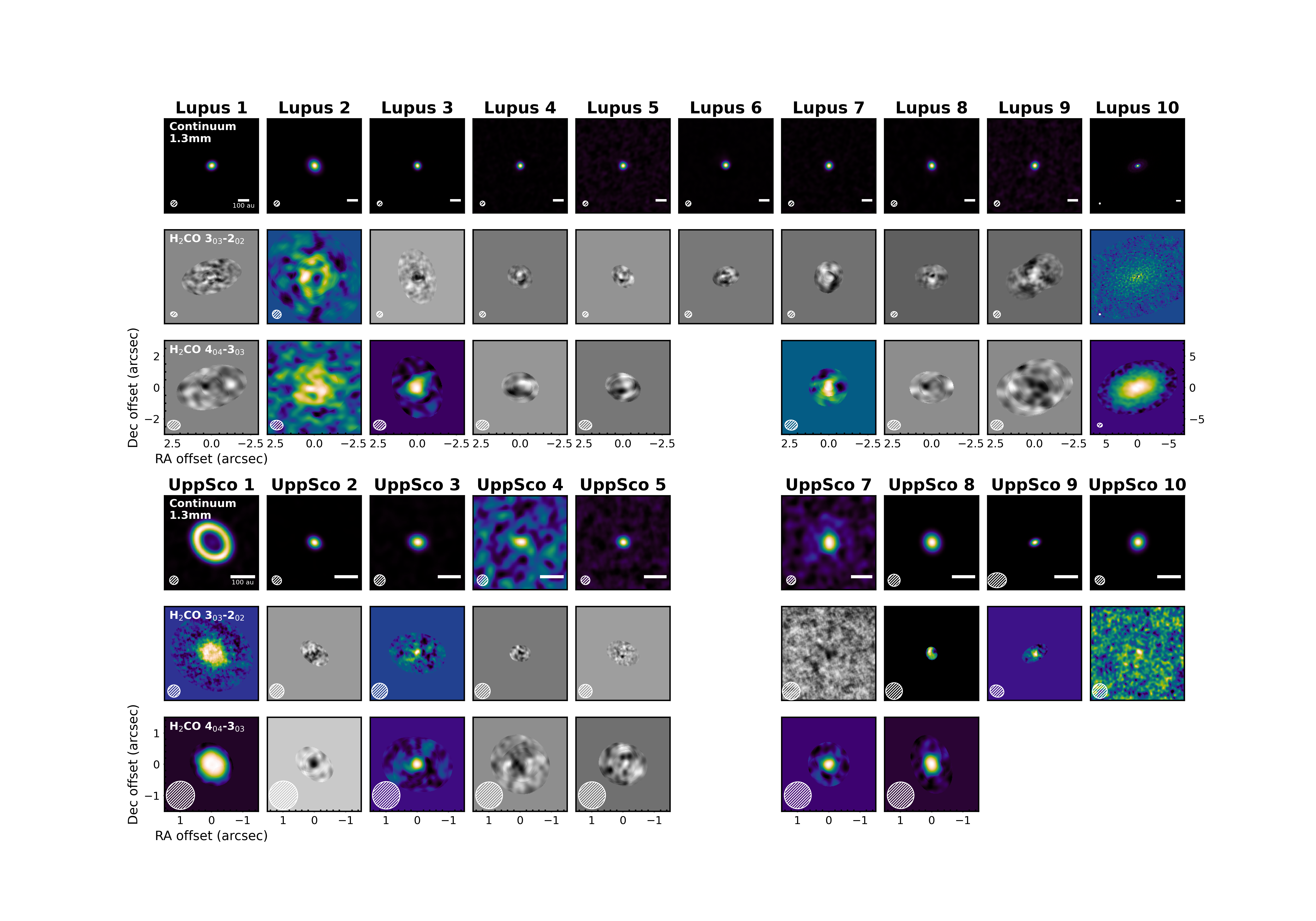}
\vspace{-0.cm}
\caption{Moment-zero maps for all of the Lupus and Upper Sco sources. The grayscale panels represent non-detections. No \htwoco~data was available for Upper Sco 6 and for the p-4-3 line for Lupus 6, Upper Sco 9, and Upper Sco 10.}
\label{fig:final_m0_all}
\vspace{-0.cm}
\end{figure*}

Figure \ref{fig:final_m0_all} shows the moment-zero maps for all AGE-PRO sources in this paper, while Figure \ref{fig:total_m0} shows only the moment-zero maps for sources with significant \htwoco~detections. 

\clearpage

\begin{figure*}[!ht]
\centering

\includegraphics[width=0.65\textwidth]{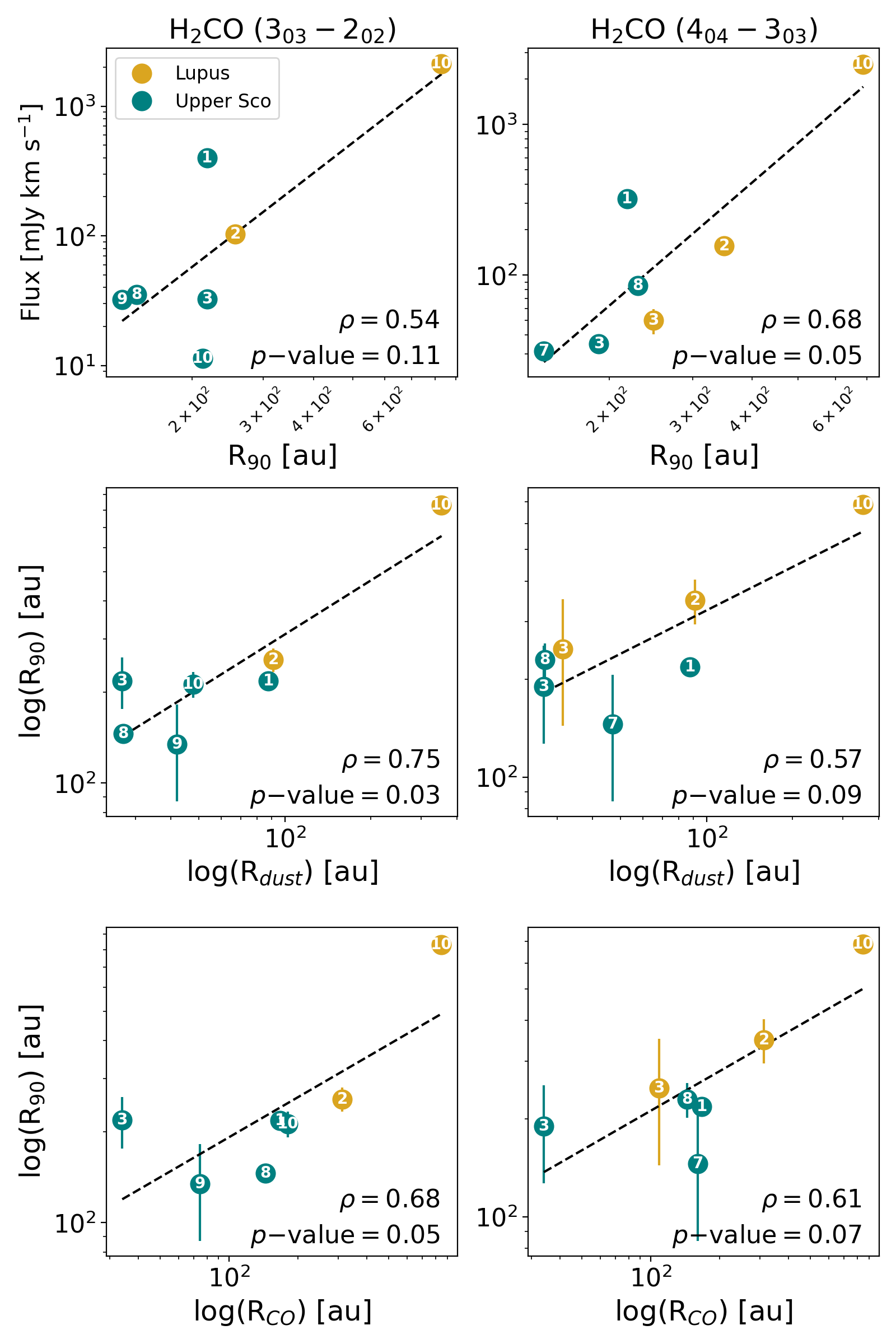}

\caption{Correlations between the \htwoco~flux and \htwoco~R$_{90}$ value, and correlations between the \htwoco~R$_{90}$ values and the dust and gas radii.}

\label{fig:r90_v_dust_radius}
\end{figure*}

Figure \ref{fig:r90_v_dust_radius} shows the correlations between the \htwoco~line flux and the R$_{90}$ value, as well as between the \htwoco~R$_{90}$ values and the dust and gas radii. The R$_{90}$ value is calculated by finding the radius at which 90$\%$ of the total flux is enclosed. The R$_{90}$ value can differ for the two \htwoco~line fluxes for a given disk. The flux of the \htwoco~p-3-2 line is found to be strongly correlated with the dust radius and borderline correlated with the gas radius, and the flux of the \htwoco~p-4-3 line is not found to be strongly correlated with the dust radius or the gas radius. However, all four plots show direct relationships that are roughly linear. We expect that the \htwoco~R$_{90}$ value will increase with the gas and dust radii, as all typically correspond with larger disk sizes.

\clearpage
\bibliography{lib}{}
\bibliographystyle{aasjournal}



\end{document}